\documentclass[twocolumn, trackchanges]{aastex631}

\usepackage{amsmath}
\usepackage{cancel}

\usepackage{graphicx}
\usepackage{amssymb}
\usepackage{bm}
\usepackage{verbatim}
\usepackage{xcolor}
\usepackage{mathtools}
\usepackage{amsfonts}
\usepackage{graphics}
\usepackage{listings}
\usepackage[normalem]{ulem}      
\usepackage{color,soul}
\def\Brms{B_{\rm rms}}

\newcommand{\EQ}{\begin{equation}}
\newcommand{\EN}{\end{equation}}
\newcommand{\EQA}{\begin{eqnarray}}
\newcommand{\ENA}{\end{eqnarray}}

\newcommand{\Eq}[1]{Eq.~(\ref{#1})}

\newcommand{\Fig}[1]{Fig.~\ref{#1}}

\newcommand{\Figs}[2]{Figs~\ref{#1} and \ref{#2}}

\newcommand{\Tab}[1]{Table~\ref{#1}}

\newcommand{\lb}{\left}
\newcommand{\rb}{\right}

\newcommand{\wt}{\tilde}
\newcommand{\p}{\partial}
\newcommand{\ti}{\textit}

\newcommand{\dd}{\mathrm{d}}        
\newcommand{\Rey}{\text{Re}}  
\newcommand{\Rm}{R_\text{m}}  
\newcommand{\Pm}{\text{Pm}}  
\newcommand{\tff}{t_{\rm ff}}

\newcommand{\cm}{\,{\rm cm}}     
\newcommand{\ccinv}{\,{\rm cm}^{-3}}     

\newcommand{\km}{\,\text{km}}    
\newcommand{\AU}{\,{\rm AU}}     
\newcommand{\psc}{\,{\rm pc}}     
\newcommand{\kpc}{\,{\rm kpc}}  

\newcommand{\s}{\,{\rm s}}      
\newcommand{\Myr}{\,{\rm Myr}} 
\newcommand{\kms}{\km\s^{-1}}    

\newcommand{\G}{\,{\rm G}}      
\newcommand{\kG}{\,{\rm kG}} 

\newcommand{\K}{\,{\rm K}}      
\newcommand{\etad}{\eta_{\rm d}} 
\newcommand{\vsat}{v_{\rm sat}} 
\newcommand{\Rmcr}{\Rm^{\text{cr}}}
\newcommand{\const}{\text{const}}
\newcommand{\rvir}{R_{\rm vir}}
\newcommand{\vvir}{V_{\rm vir}}
\newcommand{\Tvir}{T_{\rm vir}}
\newcommand{\Msun}{M_{\odot}}

\newcommand{\machsat}{\mathcal{M}_{\rm sat}}
\newcommand{\vrms}{v_{\rm rms}}
\newcommand{\Bsat}{B_{\rm sat}}
\newcommand{\btilde}{\tilde{B}_{\rm rms}}
\newcommand{\tstd}{t_{\rm sat, std}}
\newcommand{\tcol}{t_{\rm sat, col}}
\newcommand{\bfin}{B_{\rm final}}
\newcommand{\Bnu}{B_\nu}
\newcommand{\spin}{\lambda_i}

\newcommand{\lcorr}{l_{\rm corr}}

\begin{document}

\title{Collapse-accelerated small-scale dynamos in the first stars and galaxies}

\author[0009-0001-9958-7707]{Muhammed Irshad}
\affiliation{International Centre for Theoretical Sciences, 
Bangalore, 560089, India}

\email{muhammed.irshad@icts.res.in}

\author{Pallavi Bhat}
\affiliation{International Centre for Theoretical Sciences, Bangalore, 560089, India}
\email{pallavi.bhat@icts.res.in}

\begin{abstract}
Here, we explore magnetic field amplification by the small-scale dynamo
(SSD) during the formation of the first stars (FS) and first galaxies (FG).
Gravitational collapse continuously modifies the turbulent velocities and
length scales, making the SSD fundamentally different from a dynamo
operating in stationary turbulence. Previous analytical estimates of
magnetic fields during structure formation did not account for this
dynamical evolution of turbulence during collapse. In \citet{irshad},
we developed a supercomoving formulation of turbulent dynamos to address
this problem and showed that, during the kinematic stage of the SSD, the
increasing dynamo growth rate leads to a super-exponential (SE) magnetic
field amplification. Here, we show that collapse also accelerates the
otherwise slower nonlinear stage of the SSD, allowing magnetic energy to
reach progressively larger scales. 
{We then apply the supercomoving framework to models of first-star and first-galaxy formation, following the evolution of the turbulence, magnetic-field strength, and magnetic correlation scale throughout the collapse. We find that collapse-induced acceleration is particularly important when the underlying dynamo is inefficient: if the standard dynamo is already fast, it saturates before substantial collapse occurs, whereas if it is too inefficient, flux freezing dominates the amplification. Between these limits, collapse can substantially accelerate an otherwise slow dynamo, with speed-ups of \(\gtrsim10\) within the parameter space explored, enabling dynamically significant magnetic fields to develop within the available collapse time. The nonlinear evolution further drives scale-by-scale saturation toward progressively larger eddies, producing magnetic fields coherent on a non-negligible fraction of the collapsing system. Our calculations provide estimates and scaling relations for the field strength and correlation scale in first stars and galaxies, while showing how these depend on the halo, turbulence, dynamo, and collapse parameters. 
The earlier emergence of dynamically important and increasingly coherent fields can affect fragmentation, angular-momentum transport, and the subsequent evolution of the first stars and galaxies.}
\end{abstract}


\section{Introduction}
\label{sec:intro}

Magnetic fields are ubiquitous in astrophysical systems, from individual
stars to galaxies and galaxy clusters \citep{reiners_observations_2012, hull_interferometric_2019, Beck2020, Borlaff+2023}. These fields are believed to be amplified and
maintained by turbulent dynamos, which convert turbulent
kinetic energy into magnetic energy through the stretching and 
{bending} of
magnetic field lines
\citep{AxelKandu, 2016JPlPh..82f5301F,Rincon_2019,Tobias_2021,SS21}.
Depending on the correlation scale of the generated field relative to the
turbulent driving scale, dynamos are broadly classified into large-scale
dynamos (LSDs) and small-scale dynamos (SSDs). LSDs generally require
rotation, shear, or helicity and generate coherent fields on scales larger
than the turbulent driving scale. SSDs{, on the other hand} can operate, {under more generic turbulent conditions} in sufficiently
conducting plasmas \citep{Moss_Shukurov_1996, PB_Faradayrot, Sur_2019, Seta_2021_ellgal}. They amplify magnetic fields on scales below
the driving scale. In this work, we focus on SSD action during the
formation of the first stars and first galaxies.

Understanding this amplification is important because magnetic fields can
become dynamically significant {already} during gravitational collapse \citep{Turk_2012}. The rapid
development of three-dimensional magnetohydrodynamics (MHD) simulations of primordial star
formation has shown that magnetic fields can provide additional support,
modify angular-momentum transport, and affect the fragmentation of
collapsing clouds and protostellar discs
\citep{HiranoMachida2022,Sharda2020,Stacy2022, hennebelle_influence_2022, Sharda2025}.
{Observations of the present-day star-forming region Cepheus~A find that a
coherent magnetic field regulates the turbulence and organizes the infall into
ordered accretion flows rather than resisting collapse \citep{Sandhyarani2026}.}
These effects
can alter the masses and multiplicities of the first stars and potentially
their star-formation efficiency and star-formation rate \citep{krumholz_role_2019, Sadanari2024}. In the first
galaxies, magnetic fields can similarly influence the evolution of
turbulent gas, disc formation, and subsequent star formation \citep{Latif2013SSD, Latif2014SMBH, Diaz2024LW}. 
Determining
the strength and characteristic scale of the magnetic field produced
during gravitational collapse is therefore important both for
understanding the early emergence of cosmic magnetic fields {observed in high redshift galaxies \citep{farnes_faraday_2014, farnes_observed_2017, mao_detection_2017, chen_kiloparsec-scale_2024} } and for
providing physically motivated magnetic initial conditions for simulations
of primordial star and galaxy formation.

The first numerical indications of SSD action during gravitational
collapse were obtained by
\citet{sur_generation_2010,Federrath_2011,sur2012}, who found magnetic
amplification in excess of that expected from flux freezing. Resolving the
full dynamo, however, remains computationally challenging because of the
large separation between the turbulent driving and dissipative scales and overall decrease in spatial scales during compression
\citep{Axel22, Schober2026Imprints}. Analytical and semi-analytical models have therefore been
used to estimate the amplification from weak seed fields
\citep{ScleicherSSD2010,Schober_2012,2013A&A...560A..87S,Mckee1,Xu_2020NLcol}.
These approaches typically estimate the dynamo amplification assuming exponential 
growth in a prescribed turbulent state and combine it with the amplification 
due to flux freezing. {They do not, however, follow the evolution of the turbulent state itself during collapse.} 
Compression can increase the turbulent velocity even in the presence of
turbulent decay
\citep{Robertson_2012,Higashi_2021,Higashi_2022}. The dynamo therefore
operates in an evolving rather than a stationary turbulent background.

We addressed this in \citet{irshad} using the supercomoving (SC)
formulation of MHD \citep{Shandarin1980,shapiro97}. During the kinematic
stage, where the Lorentz force is negligible, we found that gravitational
collapse can progressively increase the dynamo growth rate, leading to
super-exponential magnetic-field amplification rather than the exponential
growth expected in the stationary case. At saturation, this results in
a steeper scaling of magnetic-field strength with density than that
expected from flux freezing alone. Super-exponential amplification in a decaying turbulence under gravitational collapse was  subsequently shown numerically using
the SC formulation \citep{2025AxelEvaDecay}.

In the present work, we extend this framework to follow the magnetic-field
evolution through both the kinematic and nonlinear stages and apply it to
the formation of the first stars and first galaxies. We evolve the
turbulent velocity under gravitational compression and turbulent decay,
while following the amplification of an initially weak seed field and the
progression of magnetic saturation towards larger scales. {A central aim is to determine when collapse-induced super-exponential growth can substantially accelerate an otherwise inefficient dynamo, allowing dynamically significant fields to develop within the available collapse time.} The uncertain
properties of the initial turbulence and dynamo are treated
parametrically, {allowing us to identify the regime in which this enhancement is important and} to determine how the amplification depends on
the physical conditions of the collapsing system. For first galaxies, we
follow the collapse until the formation of a rotationally supported disc,
while for first stars we follow it to densities characteristic of the
protostellar stage. We thereby estimate both the magnetic-field strength
and its characteristic correlation scale and {quantify the dynamo speed-up produced by gravitational collapse and when it becomes astrophysically important}. 

The paper is organized as follows. In Section~\ref{sec:framework}, we
introduce the supercomoving description of the collapsing turbulent system
and discuss the kinematic and nonlinear stages of magnetic amplification.
Section~\ref{sec:model} develops the model for first structures and specifies the physical
parameters governing the turbulent and magnetic evolution. In
Section~\ref{sec:results}, {we present the results from applying} the model to the formation of the first
stars and galaxies and explore the dependence on the relevant parameters.
The implications of the model are discussed in
Section~\ref{sec:impact}, followed by {a summary,} our conclusions and limitations of the model in
Section~\ref{sec:conclusions}.

\section{Analytical framework for dynamos in a collapsing medium}
\label{sec:framework}
Conventional theories of turbulent dynamos generally assume statistically stationary turbulence.
But during gravitational collapse,  compression makes the turbulent velocity and characteristic length scales time dependent. Thus the usual results for dynamos in stationary backgrounds (we call this as the `standard dynamos') cannot be applied directly.

In \cite{irshad}, we addressed this dynamical nature of turbulence and magnetic field amplification due to turbulent dynamos in a collapsing cloud by assuming a homologous collapse. There the physical gas velocity is decomposed as
\[
\bm{u} = \bm{v} + H\bm{r}, \quad H = \frac{\dd\ln{a}}{\dd t},
\] 
where $\bm r$ is the physical position, $\bm v$ is the peculiar velocity, and \(H\bm r\) is the background contraction velocity with $H < 0$ for collapse. The scale factor $a(t)$ specifies the size of the collapsing cloud.

Further simplifications in the 
MHD 
equations are possible using the \textit{supercomoving variables} introduced by \citep{Shandarin1980, shapiro97}
\begin{gather}\label{eq:scdefn}
\tilde{\bm{r}} = \bm{r}/a,\quad \dd \tilde{t} = \dd t/{a^{2}},\quad \wt{\rho}(\wt t) = a^{3} \rho(t),\notag\\
\wt{\bm{v}}(\wt t) = a\bm{v}(t), \quad \wt{p}(\wt t) = a^{5}p(t),\quad \wt{\bm{B}}(\wt t) = a^{2}\bm{B}(t),
\end{gather}
where $\rho$ is the mass density, $p$ is the gas pressure 
and $\bm{B}$ is the magnetic field.  
Note that we use the supercomoving time $\wt{t}$ instead of physical time $t$ in the analysis.
These transformations remove the homogeneous contraction and factor out the corresponding compression of the density. Importantly, $\wt{\bm{v}}$ accounts for the increase in peculiar velocity $v \propto 1/a$ in a homologous collapse and $\wt{\bm{B}}$ factors out the amplification in magnetic field $B\propto \rho^{2/3}$ due to flux freezing from overall spherical collapse. 
The resulting supercomoving MHD (SC-MHD) equations are 
\begin{gather}
 \frac{\p\wt{\rho}}{\p\wt t}+ \wt{\nabla}\cdot\lb(\wt\rho\,\wt{\bm{v}}\rb) 
= 0, \label{eq2a:mhd_sum} 
\\[3pt]
 \frac{\p \wt{\bm{v}}}{\p \wt{t}} + (\wt{\bm{v}}\cdot\wt{\nabla})\wt{\bm{v}}
 = - \frac{\wt{\nabla} \wt{p}}{\wt{\rho}} +\wt{a}\frac{(\wt{\nabla}\times\wt{\bm{B}})\times\wt{\bm{B}}}{4\pi\wt{\rho}} + \nu \wt{\nabla}^2 \wt{\bm{v}},   \label{eq2b:mhd_sum}
 \\[3pt]
\frac{\p \wt{\bm{B}}}{\p \wt{t}} = \wt{\nabla}\times(\wt{\bm{v}}\times\wt{\bm{B}})+\eta\wt{\nabla}^{2}\wt{\bm{B}},\quad \wt{\nabla}\cdot\wt{\bm{B}} = 0,\label{eq2c:mhd_sum}
\end{gather}
where \(\wt a(\wt t)\equiv a(t)\) and $\wt{p} = \wt{p}(\wt{\rho})$ to close the equations. 
Note that SC-MHD equations differ from the standard MHD equations only by $\wt{a}(\wt t)$ in the coefficient of Lorentz force. To simplify the problem
we assume the compressibility to be small in the supercomoving variables and replaced the continuity equation with the incompressibility condition.

In the kinematic regime, where the Lorentz force is negligible, the induction equation for \(\tilde{\bm B}\) has the standard MHD form. Consequently, standard kinematic dynamo results can be extended to the collapsing system when expressed in supercomoving variables. 
The modified Lorentz force weakens with compression since $\wt a < 1$ and this is important during the nonlinear saturation of dynamos.

We approximate the background evolution by pressureless free fall. The scale factor is then determined implicitly by
\begin{equation}\label{eq:scale}
    \arctan\lb(\sqrt{\frac{1-a}{a}}\rb) + \sqrt{a(1-a)} = st,
\end{equation}
where $s = \pi / 2\tff$, and
\[
\tff=\sqrt{\frac{3\pi}{32G\rho_0}}
\]
is the free-fall time defined using the initial mean density ($\rho_0$), including both baryons and dark matter (DM). A useful simple fit for the above trajectory is \citep{Girichidis}
\begin{equation}\label{eq:scale_fit}
  a(t) = \left(1 - \frac{t^{2}}{\tff^2}\right)^{2/3}.
\end{equation}
In terms of supercomoving time, the scale factor is
\begin{equation}\label{eq:scale2}
    \wt{a}(\wt{t}) = \left(1 + \frac{s^2 \wt{t}^2}{4}\right)^{-1}.
\end{equation}
Equivalently,
\begin{equation}\label{eq:sctime}
    \tilde t = \frac{4\tff}{\pi} \sqrt{\frac{1-a}{a}}.
\end{equation}

{For the analytical development in this section, we assume maintained turbulence in supercomoving variables in order to isolate the effect of collapse on the dynamo. In the first-star and first-galaxy models of Section~\ref{sec:model}, we relax this assumption and allow the supercomoving turbulence to decay while the physical turbulent velocity evolves under gravitational compression.}

\subsection{Kinematic stage}
In the kinematic dynamo theory of statistically stationary turbulence, the magnetic field grows exponentially in time. 
Since the induction equation in the supercomoving variables has the same form as in standard MHD when the Lorentz force is negligible, the supercomoving magnetic field grows exponentially in supercomoving time,
$\wt{B} \propto \exp{\left(\wt\gamma\wt t\right)}$. In physical variables, this corresponds to a \textit{super-exponential} growth of the magnetic field
\begin{equation}\label{eq:kcoldynamo3}
B = B_0\ \xi^{2/3}\ \exp\left(\int_0^t \frac{\wt{\gamma}\,\dd t'}{a^2(t')} \right) ,
\end{equation}
where $\xi = \rho(t) / \rho_0 =a^{-3}(t)$ is the compression factor, $B_0$ is the seed magnetic field, and $\wt\gamma$ is the dynamo growth rate. The factor \(\xi^{2/3}\) corresponds to amplification by flux freezing, while the exponential factor describes the dynamo mechanism. 

The kinematic stage ends once the magnetic energy reaches equipartition with the kinetic energy of the fastest turbulent eddies of scale \(\wt l\)
\begin{align}\label{eq:sat1}
    \wt{a}\,\frac{\wt{B}^2}{8\pi} \simeq \frac{1}{2}\wt{\rho}\wt{v}_l^2,
\end{align}
where $\wt{v}_l$ is the velocity at scale $\wt l$. The factor $\wt{a}$ in the equipartition follows from the weakened Lorentz force in the SC-MHD equations. 

\subsection{Nonlinear regime}
After the kinematic stage, the fastest eddies are no longer able to stretch the magnetic field efficiently because of Lorentz force back reaction. But magnetic amplification continues through progressively larger and more energetic eddies in the inertial range. The magnetic correlation scale therefore grows through scale by scale saturation. And in stationary subsonic turbulence, the nonlinear stage is shown to have a linear growth of magnetic energy both theoretically and numerically \citep{Schekochihin_2002, Beresnyak2012, Kriel2026}.

We now derive the corresponding behavior in a collapsing cloud, assuming maintained turbulence in supercomoving variables. The supercomoving magnetic energy $\wt W = \wt{B}^2 / 8\pi$, evolves in the high conductivity limit according to \citep{Schleicher2013}
\begin{align}\label{eq:nl1}
    \frac{d\wt W}{d\wt t} = \wt\Gamma \wt W,
\end{align}
where $\wt{\Gamma} = \wt{v}_l / \wt l$ is the turnover rate of the fastest effective eddy. For a turbulent spectrum satisfying
$\wt{v}_l \propto \wt{l}^{\vartheta}$, the growth rate can be written as
\begin{align}\label{eq:nl2}
    \wt{\Gamma} = \frac{\wt{v}_0^{1/\vartheta}}{\wt{l}_0} \; \wt{v}_l^{(\vartheta - 1)/\vartheta},
\end{align}
where \(\wt v_0\) and \(\wt l_0\) are the supercomoving velocity and length scale {at the energy-containing (integral) scale} of the maintained supercomoving turbulence. Assuming the nonlinear stage proceeds through scale by scale equipartition, we substitute Eq.~\eqref{eq:sat1} in Eq.~\eqref{eq:nl2}, 
to express $\wt\Gamma$ in terms of $\wt{W}$, 
\begin{align}\label{eq:nl3}
    \wt{\Gamma} = \frac{\wt{v}_0^{1/\vartheta}}{\wt{l}_0} \; \left(\frac{2}{\langle\wt{\rho}\rangle}\right)^{\frac{(\vartheta-1)}{2\vartheta}}\  \left(\wt{a}\wt{W}\right)^{\frac{(\vartheta-1)}{2\vartheta}}  =  C_0 \left(\wt{a}\wt{W}\right)^{\frac{(\vartheta-1)}{2\vartheta}}
\end{align}
where $C_0$ is a constant. Substitute Eq.~\eqref{eq:nl3} in Eq.~\eqref{eq:nl1} to obtain the evolution of magnetic energy density in the nonlinear stage
\begin{align}\label{eq:nl4}
    \frac{d\wt W}{d\wt t} = C_0 \wt{W} \left(\wt{a}\wt{W}\right)^{(\vartheta-1)/2\vartheta}.
\end{align}
Using Eq.~\eqref{eq:scale}, $\dd \wt t = \dd t / a^2 = -s^{-1} a^{-3/2} (1 - a)^{-1/2}\,\dd a$ and substitute this to get
\begin{align}\label{eq:wtilde1}
    \int \wt W^{\lb(1-3\vartheta\rb) / {2\vartheta}}\,\dd\wt W &= -\frac{C_0}{s}\int (1 - a)^{-1/2}\, a^{-(1 + 2\vartheta) / 2\vartheta}\, \dd a \notag
    \\
    &= -\frac{C_0}{s}\, B\lb(a; -\frac{1}{2\vartheta}, \, \frac{1}{2}\rb) + w_1,
\end{align}
where \(B(a;p,q)\) is the incomplete beta function and \(w_1\) is an integration constant fixed by the magnetic energy at the end of the kinematic stage.

A particularly simple closed form is derived from Eq.~\eqref{eq:wtilde1} for $\vartheta = 1/3$ relevant for the subsonic turbulence. Using $x^2 = (1-a)/a$ and Eq.~\eqref{eq:sctime} to obtain
\[
\int (1 - a)^{-1/2}\, a^{-5 / 2}\, \dd a = -\lb(s\wt t + \frac{s^3}{12}\, \wt t^{\,3}\rb).
\]
Consequently, the supercomoving magnetic energy density grows as
\begin{gather}\label{eq:nl5}
    \wt{W} = K \left(\wt{t} + \frac{s^2 \wt{t}^{\,3}}{12}\right) + w_1,
\end{gather}
where $K = \wt{\rho} \wt{v}_0^3 / 2 \wt{l}_0$. An even simpler derivation follows from Eqs~\eqref{eq:sat1} and \eqref{eq:nl1} 
\begin{align}
    \frac{\dd\wt W}{\dd \wt t} &= \wt\Gamma\wt W = \frac{\wt\rho\wt{v}_l^3}{2\wt l}\ \frac{1}{\wt{a}}.
\end{align}
Assuming constant energy flux we integrate this to obtain Eq.~\eqref{eq:nl5}.

Note that unlike the linear growth of magnetic energy in the nonlinear stage of a stationary dynamo,
the magnetic energy grows \ti{super-linearly} in supercomoving time in a collapsing background. This is due to weakening of the SC-Lorentz force. Moreover, in the physical time this is an even faster growth because of the shrinking in dynamical time scales during collapse.
This boost to the nonlinear dynamo can substantially shorten the slow scale by scale saturation stage and produce stronger magnetic fields coherent on larger scales before the collapse ends.

\subsection{Saturation}
At the end of scale-by-scale saturation, the magnetic energy reaches equipartition with a fraction \(f_B\) of the turbulent kinetic energy at the injection scale \(\wt l_0\)
\begin{align}
\wt a\frac{\wt B^2}{8\pi}
\simeq
f_B\,\frac{1}{2}\wt\rho\,\wt v_0^2,
\end{align}
where \(f_B\) is the magnetic saturation efficiency. Transforming back to physical variables gives
\[
\Bsat
=
\sqrt{4\pi f_B\,\rho v^2}.
\]
For the adiabatic heating of velocity $v \propto a^{-1} \propto \rho^{1/3}$, the saturated magnetic field scales with density as $\Bsat \propto \rho^{5/6}$.

\section{Model for turbulence and magnetic field evolution in first stars and galaxies}
\label{sec:model}

Primordial density perturbations generated during inflation provided the seeds for cosmic structure formation \citep{Weinberg:2008}. Under gravitational instability, overdense regions accrete matter from their surroundings while initially expanding with the Hubble expansion. Once their self gravity became sufficiently strong, they decouple from the cosmic expansion, collapse nonlinearly, and virialize into dark matter (DM) haloes containing baryons \citep{Peebles}.

The baryons subsequently evolve within the gravitational potential wells of these DM haloes. During infall and virialization, a fraction of the gravitational binding energy is converted into thermal energy through gas collisions and accretion shocks. The gas is thereby heated to approximately the virial temperature  \citep{Barkana_2001, Loeb_Furlanetto_2013}
\begin{align}\label{eq:Tvir}
    \Tvir = \frac{\mu m_H \vvir^2}{2k_B},
\end{align}
where $\mu$ is the mean molecular weight 
\footnote{{We assume the collapsing primordial gas to remain predominantly neutral  and adopt \(\mu=1.22\), appropriate for a neutral primordial H--He mixture \citep{Greif_Firstgalaxy2008, 2013ASSL..396..103G}.}}, $m_H$ is the mass of hydrogen and $k_B$ is the Boltzmann constant. The virial velocity
\begin{align}\label{eq:vvir}
    \vvir = \sqrt{\frac{GM}{\rvir}},
\end{align}
sets the characteristic speed associated with a DM halo of total mass \(M\). Its virial radius is
\begin{align}\label{eq:rvir}
R_{\rm vir} = \lb(
\frac{3M}{4\pi \Delta_{\rm vir}\rho_{\rm bg}(z)}\rb)^{1/3}.
\end{align}
Here \(\rho_{\rm bg}(z)\) is the mean cosmological matter density
\begin{align}
\rho_{\rm bg}(z)
=
\Omega_m\rho_{\rm cr}(1+z)^3
=
\frac{3H_0^2\Omega_m}{8\pi G}(1+z)^3,
\end{align}
where \(H_0\) is the present day Hubble parameter, \(\Omega_m\) is the matter-density parameter today, and \(z\) is the redshift. We adopt the top-hat model, and take virial overdensity $\Delta_{\rm vir}\simeq 18\pi^2$.
Typical values of $\Tvir,\, \vvir,\, \rvir$ are given in Table.~\ref{tab:typical_vals} for the First stars and First galaxies we are interested in.

Virialized gas is initially supported against further gravitational contraction by its thermal pressure. Continued collapse therefore requires radiative cooling. In low mass minihaloes, with characteristic masses \(M\lesssim10^6\,M_\odot\), cooling is dominated by molecular hydrogen, \(\mathrm{H}_2\). In more massive atomic cooling haloes, atomic hydrogen and helium line emission provide the dominant cooling channels \citep{Abel_model_primgas,Galli_Palla_98,Glover_2005review}. Cooling reduces the pressure support and allows the gas to contract to much higher densities, creating the conditions for the formation of the first stars in minihaloes and the first galaxies in atomic-cooling haloes \citep{2009Natur.459...49B,1stgalaxies2011}.

\subsection{Seed field}
A turbulent dynamo requires a nonzero seed magnetic field to act on. Possible seed field mechanisms can be broadly divided into primordial and astrophysical sources. Primordial fields may be generated during inflation or cosmological phase transitions, although both their amplitude and coherence length at the epoch of halo formation remain highly model dependent \citep{2016RPPh...79g6901S}. Astrophysical seed fields can instead be generated after recombination through battery processes, shocks, and plasma instabilities during structure formation. 

In this work, we adopt the Biermann battery mechanism at the oblique shock fronts formed during the process of formation of astrophysical objects, where magnetic fields could be generated if the pressure and density gradients are misaligned \citep{1989ApJ...342..650P, Xu_2008}. Mathematically this is analogous to the baroclinic term responsible for vorticity production. Using the similarity between the vorticity ($\bm\omega$) and induction equations, \cite{Mckee1} gives the relation
\begin{align}\label{eq:seed}
    \bm{B}_0 = -\lb[\frac{\mu m_H\, c}{e(1+x_e)}\rb]\, \bm\omega,
\end{align}
where $c$ is the speed of light, $e$ the charge of electron, the vorticity is taken to be $|\omega| \sim \vvir / \rvir$ and $x_e = n_e /n$ is the ionization fraction with \(n_e\) and \(n\) denoting the electron and total number densities, respectively.

We use Eq.~\eqref{eq:seed} to estimate the fiducial seed field given the gas composition and halo properties of each model. In the extended parameter study, we then  vary \(B_0\) over \(10^{-20}\) - \(10^{-10}\,\G\). This range {can represent} weak to stronger fields from {either Biermann battery or primordial magnetogenesis} to determine how sensitively, dynamo during the collapse,  depends on the seed field.

\subsection{Generation of turbulence}
The accretion of baryons into a DM halo converts part of the released gravitational energy into turbulent motions through accretion shocks, collisions between converging gas streams, and baroclinic vorticity generation \citep{Kl_Hen2010Accretion}. Such turbulent motions are seen in simulations of both first star and first galaxy formation \citep{Greif_Firstgalaxy2008, Grief2012}. Of particular importance for the small-scale dynamo is the vortical component of the flow. Vorticity can be generated downstream of curved shocks even when the upstream flow is initially irrotational. For a strong shock, the characteristic vorticity is \(|\bm\omega|\simeq{2U} / {R}\),
where \(U\) is the upstream velocity and \(R\) is the smaller of the two principal curvature radii of the shock front \citep{Doroshkevich1970,SS21}. The resulting vortical motions interact nonlinearly and establish a turbulent cascade within a few eddy turnover times.

Since the detailed turbulent state depends on the assembly history of each halo, we parametrize the initial physical integral scale and velocity as
\begin{align}
    l_{0i} = f_l\rvir, \quad v_{0i} = f_v c_s, 
\end{align}
where $f_l$ and $f_v$ are dimensionless parameters constrained by numerical simulations. The sound speed is
\begin{equation}\label{eq:cs}
    c_s = \sqrt{\frac{\gamma k_B T}{\mu m_H}},
\end{equation}
where the temperature is set to \(T=\Tvir\) assuming approximately isothermal evolution ($\gamma = 1$) during the collapse of primordial gas. Typical sound speeds are given in Table.~\ref{tab:typical_vals}.

\subsection{Evolution of turbulence}
In the absence of sustained driving, the turbulence generated at virial shocks decays through the turbulent cascade. However,
during gravitational collapse this decay {can be} counteracted by the compression of the gas, which injects energy into the physical turbulent motions \citep{Robertson_2012, Hennebelle2021, Higashi_2021,Higashi_2022}. We model this using the decaying supercomoving turbulence developed in the supplementary material of \citet{irshad}.

The increase in physical velocity due to compression is already contained in the definition of the supercomoving velocity. We therefore describe the supercomoving turbulence as freely decaying  
\begin{equation}\label{eq:dec1}
    \frac{\dd \wt{v}^2}{\dd \wt t} = -{\etad}\; \frac{\wt{v}^3}{\wt l},
\end{equation}
where $\etad$ is the dissipation factor. Based on the studies of \cite{1999ApJ...524..169M, Higashi_2022} we take $\etad = 0.42$. {Assuming 
a peaked velocity spectrum, we can constrain} the evolution of SC integral scale {by considering an invariant quantity which can be expressed as}
\begin{equation}\label{eq:dec2}
	\wt{v}_0^\alpha \wt{l}_0 = \const = {\wt{v}_{0i}^\alpha \wt{l}_{0i} = \psi_0}.
\end{equation}
Here the subscript \(i\) denotes the initial value. 
The choices \(\alpha=2/5\) and \(\alpha=2/3\) correspond to the Loitsyansky and Saffman invariants respectively \citep{Davidson2015}. We take $\alpha = 2/3$, as it {can be more} appropriate for compressible flows (see section 14.3.2 of \cite{SS21} on decaying turbulence). 

Solving Eqs~\eqref{eq:dec1} and \eqref{eq:dec2} together with the $\wt a$ model in Eq.~\eqref{eq:scale2} yields
\begin{align}\label{eq:dec3}
    \tilde{v}_0 = \tilde{v}_{0i} \lb[1 + (1+\alpha)\chi \sqrt{\frac{1-a}{a}}\ \rb]^{-1/(1+\alpha)},
\end{align}
where $\chi = \eta_d / s \tau_L$, {with $s = {\pi}/{2 t_{\rm ff}}$ (see \Eq{eq:scale})} and $\tau_L = \tilde{l}_{0i} / \tilde{v}_{0i}$, is a dimensionless parameter proportional to the ratio of the free fall time to the initial integral scale turn over time. Asymptotically, for $a\ll 1$, from Eq.~\eqref{eq:dec3} we get the scaling $v_0 = \tilde{v}_0 / a \propto \rho^{(1+2\alpha) / \lb[6(1+\alpha)\rb]}$. For $\alpha=2/3$ of our interest, $v_0 \propto \rho^{7/30}$. 

Note that \(\wt v_0\) decreases during collapse and the decay is stronger for larger \(\chi\). 
The  Kolmogorov scaling $\tilde{v}_l \propto \tilde{l}^{1/3}$ captures the evolution of inertial range. 
Even in supersonic turbulence, the Kolmogorov spectrum is a useful approximation on most of the scales where $v_l < c_s$ (sonic scale) \citep{Federrath2021}. We therefore use this scaling throughout the present model.

Although the SC turbulence decays, corresponding physical velocity $v_0=\tilde{v}_0/a$ increases due to gravitational compression. But this increase cannot continue indefinitely. Once the turbulent Mach number becomes of order unity, shocks and {dilatational motions provide additional channels for} dissipation \citep{Alam2026, Troccoli2026}.
Motivated by these and by {the saturation of turbulent velocity in} \cite{Higashi_2022}, we impose a {limiting} velocity
$\vsat = \machsat\,c_s$, 
where $\machsat$ is {a free parameter specifying the} {limiting} Mach number {and its value is informed by simulations of first star and first galaxy formation}.

\begin{figure}
    \centering
    \includegraphics[width=\linewidth]{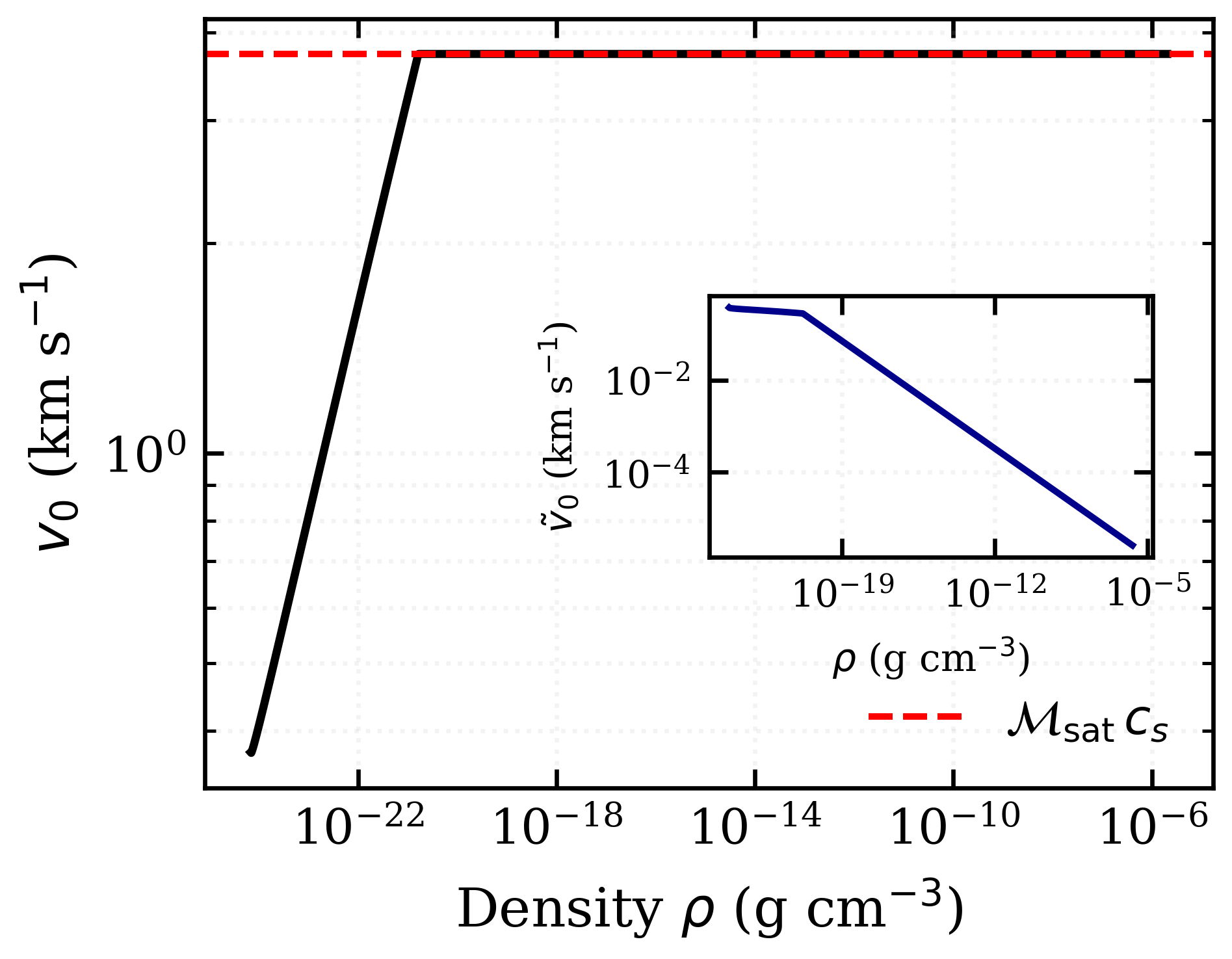}
    \caption{Evolution of the physical integral scale velocity \(v_0\) with increasing density for the fiducial first star model. The red dashed line denotes the imposed saturation limit \(v_{\rm sat}=\machsat c_s\). The inset shows the corresponding decay of the supercomoving velocity \(\wt v_0\). }
    \label{fig:vel_SCvel_FS}
\end{figure}

Figure~\ref{fig:vel_SCvel_FS} illustrates the velocity evolution for the fiducial first star model. The inset shows the free decay of the supercomoving velocity. In physical variables, gravitational compression compensates for this decay and drives \(v_0\) toward the imposed saturation velocity.

\begin{figure*}[t]
    \centering
     \includegraphics[width=\linewidth]{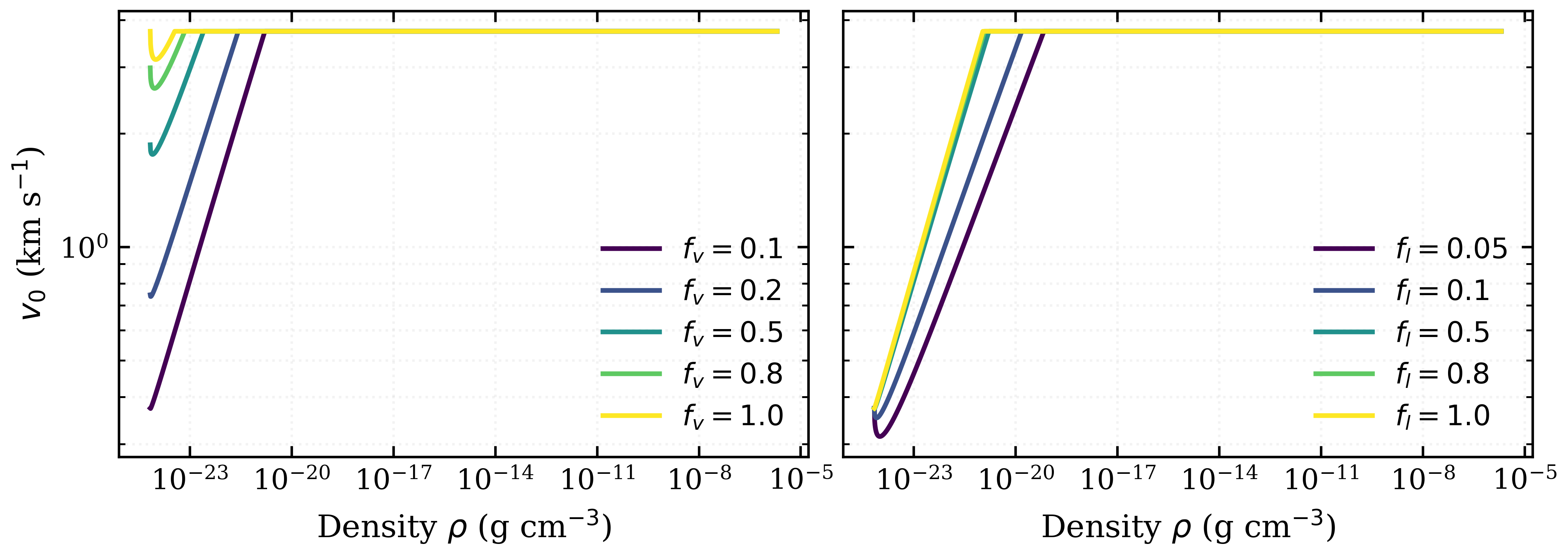} 
    \caption{Physical integral scale velocity as a function of density for different values of \(f_v\) (left) and \(f_l\) (right). Larger \(f_v\) or smaller \(f_l\) increases \(\chi\), producing stronger decay in supercomoving variables before compression raises the physical velocity to \(v_{\rm sat}\).}
    \label{fig:vel_fvfl}
\end{figure*}

\begin{figure*}[t]
    \centering
     \includegraphics[width=\linewidth]{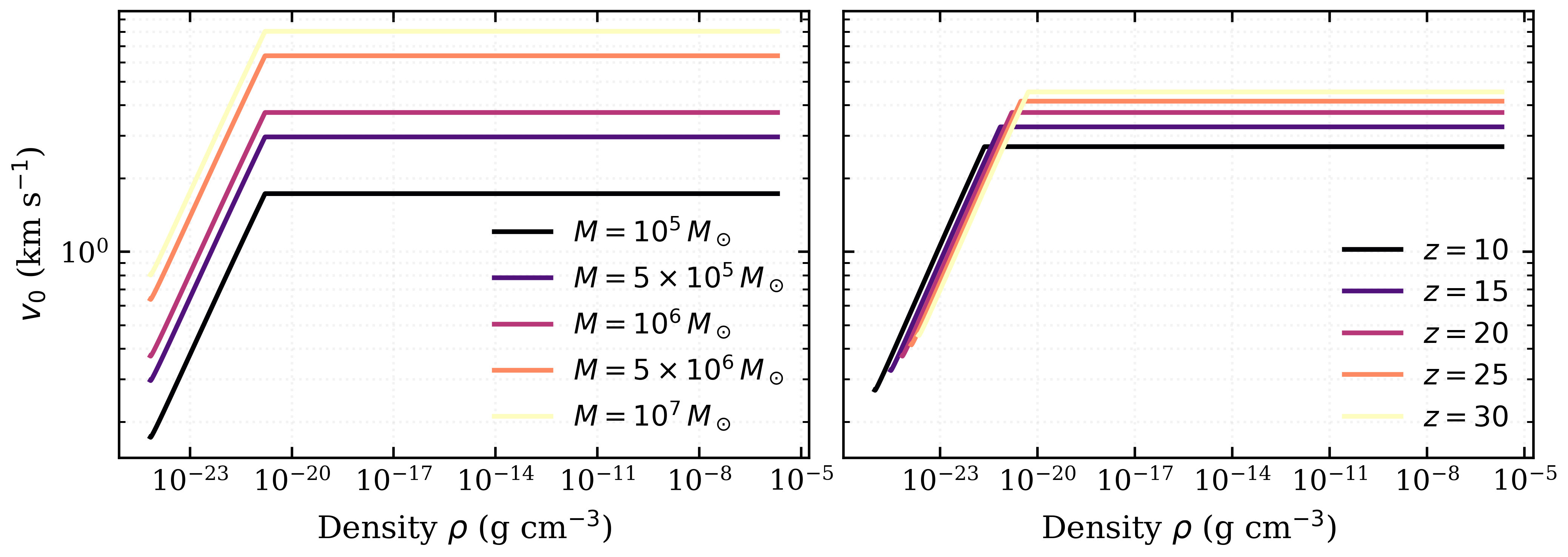} 
    \caption{Same as Fig.~\ref{fig:vel_fvfl}, but for different halo masses \(M\) (left) and formation redshifts \(z\) (right). Increasing \(M\) or \(z\) raises the sound speed and the initial velocity through the virial temperature. The relative strength of supercomoving decay remains unchanged because \(\chi\) is independent of \(M\) and \(z\) in the present model.}
    \label{fig:vel_Mz}
\end{figure*}

The turbulence evolution depends mainly on the following parameters:
\begin{enumerate}
    \item \textit{Initial velocity and scale.}
    Increasing \(f_v\) increases the initial velocity \(\wt v_{0i}\), whereas decreasing \(f_l\) reduces the initial integral scale. Since \(\chi\propto \tff / \tau_L \propto f_v / f_l,\) increasing \(f_v\) or decreasing \(f_l\) increases the {initial} turbulent decay {in $v_0$}. 
    Fig.~\ref{fig:vel_fvfl} shows the resulting variation {in $v_0$ with $f_v$ and $f_l$}. 
    \item \textit{Halo mass ($M$) and formation redshift ($z$).}
    They set the virial temperature, sound speed, and virial radius. With redshift the free-fall time goes as $\tff \propto (1 + z)^{-3/2}$.
Since \(\tau_L\propto {\rvir} / {c_s}\propto(1+z)^{-3/2},\)
the parameter \(\chi\) and the decay rate is independent of both \(M\) and \(z\). By contrast, the initial velocity scales as \(\wt v_{0i}\propto c_s \propto M^{1/3}(1+z)^{1/2}.\)
Thus increasing either halo mass or formation redshift raises the initial velocity without changing the relative strength of turbulent decay. This behavior is illustrated in Fig.~\ref{fig:vel_Mz}.
\end{enumerate} 

The corresponding supercomoving integral scale before the velocity cap follows from  Eq.~\eqref{eq:dec2}
\begin{align}\label{eq:decl0}
    \tilde{l}_0 = \tilde{l}_{0i} \lb[1 + (1+\alpha)\chi \sqrt{\frac{1-a}{a}}\rb]^{\alpha/(1+\alpha)},
\end{align}
and the physical integral scale is $l_0 = a \tilde{l}_0$. A larger $\chi$ produces stronger supercomoving decay and therefore a larger $\wt{l}_0$, {because turbulent decay dissipates the smaller-scale motions first, so that the remaining turbulent energy is progressively dominated by larger scales}. 
Once the physical velocity saturates, the
self similar decay solution from which Eq.~\eqref{eq:dec2} follows no longer
applies. We therefore {no longer assume the invariant relation to hold beyond this point}. {Instead}, the
integral scale is {taken to be} simply advected with the overall collapse
\begin{align}\label{eq:dec4}
    l_0 = l_0(a_{\vsat})\,\frac{a}{a_{\vsat}}  = \mathcal{L}_{0,\rm sat}\, a,
\end{align}
where $a_{\vsat}$ is the scale factor at the time of velocity saturation $v_0 = \vsat$ and we defined \(\mathcal{L}_{0,\rm sat} = l_0(a_{\vsat}) / a_{\vsat}\).

\begin{table}[t]
\centering
\caption{Representative values for the fiducial first-star ($M_h=10^6\,M_\odot$, $z=20$) and first-galaxy ($M_h=10^8\,M_\odot$, $z=10$) haloes {with mean molecular weight of $\mu=1.22$} considered in this work. The values are approximate and are computed using the relations described in the text. Specifically seed field $B_0$ is computed from Eq.~\eqref{eq:seed} with $\omega = \vvir / \rvir$. The kinetic and magnetic Reynolds numbers are defined by \(\Rey=v_0l_0/\nu\) and \(\Rm = v_0l_0 / \eta\), respectively. }
\label{tab:typical_vals}
\begin{tabular}{lcc}
\hline
Parameter & First star & First galaxy \\
\hline
$M_h\, (M_\odot)$                 & $10^6$ & $10^8$ \\
$z$                   & $20$ & $10$ \\
$R_{\rm vir}\, (\kpc)$         & $0.15$ & $1.37$ \\
$V_{\rm vir}\,(\kms)$         & $5.3$ & $17.7$ \\
$T_{\rm vir}\,(\K)$         & $2.1\times10^3$ & $2.3\times10^4$ \\
$c_s\,(\kms)$                & $3.73$ & $12.53$ \\
$\Rey$    & $1.05\times 10^4$      & 1.3$\times 10^4$   \\
$\Rm$     &  $1.6\times 10^{16}$  & 7.67$\times 10^{17}$  \\
$\tff\,(\Myr)$   &    31.85      & 84.0 \\
$\omega$ (in $\rm{s}^{-1}$)   &    $1.1\times10^{-15}$     & $4.2\times10^{-16}$  \\
$B_0\,(\G)$                  & $1.4\times10^{-19}$ & $5.25\times10^{-20}$ \\
\hline
\end{tabular}
\end{table}

\subsection{Reynolds numbers}
To understand the properties of turbulence and the turbulent dynamos, it is useful to estimate the kinetic and magnetic Reynolds numbers \(\Rey = v_0 l_0 / \nu,\quad\, \Rm = v_0 l_0 / \eta,\) where \(\nu\) is the kinematic viscosity and \(\eta={c^2}/{4\pi\sigma}\) is the magnetic diffusivity associated with the Ohmic conductivity \(\sigma\).
From eqs~\eqref{eq:dec3}, \eqref{eq:decl0}, and \eqref{eq:dec4}, we get the numerator
\begin{equation}\label{eq:v0l0}
 v_0 l_0 = 
\begin{cases}
    \tilde{v}_{0i}\,\tilde{l}_{0i}\,\lb[1 + (1+\alpha)\chi \sqrt{\frac{1-a}{a}}\,\rb]^{\frac{(\alpha - 1)}{(\alpha + 1)}},   & \text{if } v_0 < \vsat \\[5pt]
    \vsat \,\mathcal{L}_{0,\rm sat}\, a\ ,  \qquad \text{otherwise}.
\end{cases}
\end{equation}

The primordial gas {in first stars and galaxies} is partially ionized and its viscosity is {assumed to be} dominated by neutral particle collisions. So we use the relevant kinematic viscosity (see \S{2.6.4} on Viscosity of a Partially Ionized Plasma in \citet{SS21})
\begin{align}\label{eq:nu}
    \nu \simeq 2.0\times 10^{20}\,\cm^2\,\s^{-1}\,\lb(\frac{n_n}{1\,\ccinv}\rb)^{-1}\,T_3^{1/2},
\end{align}
where $n_n$ is the number density of neutrals and $T_3 = T / 10^3 \K$. This viscosity is larger than that of a fully ionized plasma because neutral interactions are weaker than Coulomb interactions and therefore allow longer mean free paths. As the gas density increases during collapse $n_n$ increases, $\nu$ decreases, and $\Rey$ rises. 

The effective magnetic diffusivity given in the appendix of \cite{Mckee1} is \cite[see also][]{PintoGalli08}
\begin{align}\label{eq:eta}
\eta={}&
6.5\times10^8\,T_3^{-3/2}
\\
&+
4.0\times10^8\,T_3^{0.6}
x_{e,-4}^{-1}
\exp\left(-0.43\sqrt{T_3}\right)
\quad \cm^2\,\s^{-1},\nonumber
\end{align}
where $x_{e,\,-4} = x_e / 10^{-4}$ is the normalized ionization fraction and we adopt $x_e=10^{-4}$.
For fixed \(T\) and \(x_e\), \(\eta\) is independent of density {and remains constant during collapse. The evolution of \(\Rm\) therefore arises solely through the} Eq.~\eqref{eq:v0l0}. 
Whereas the strong density dependence of \(\nu\) causes a much larger increase in \(\Rey\).

\begin{figure}
    \centering
    \includegraphics[width=\linewidth]{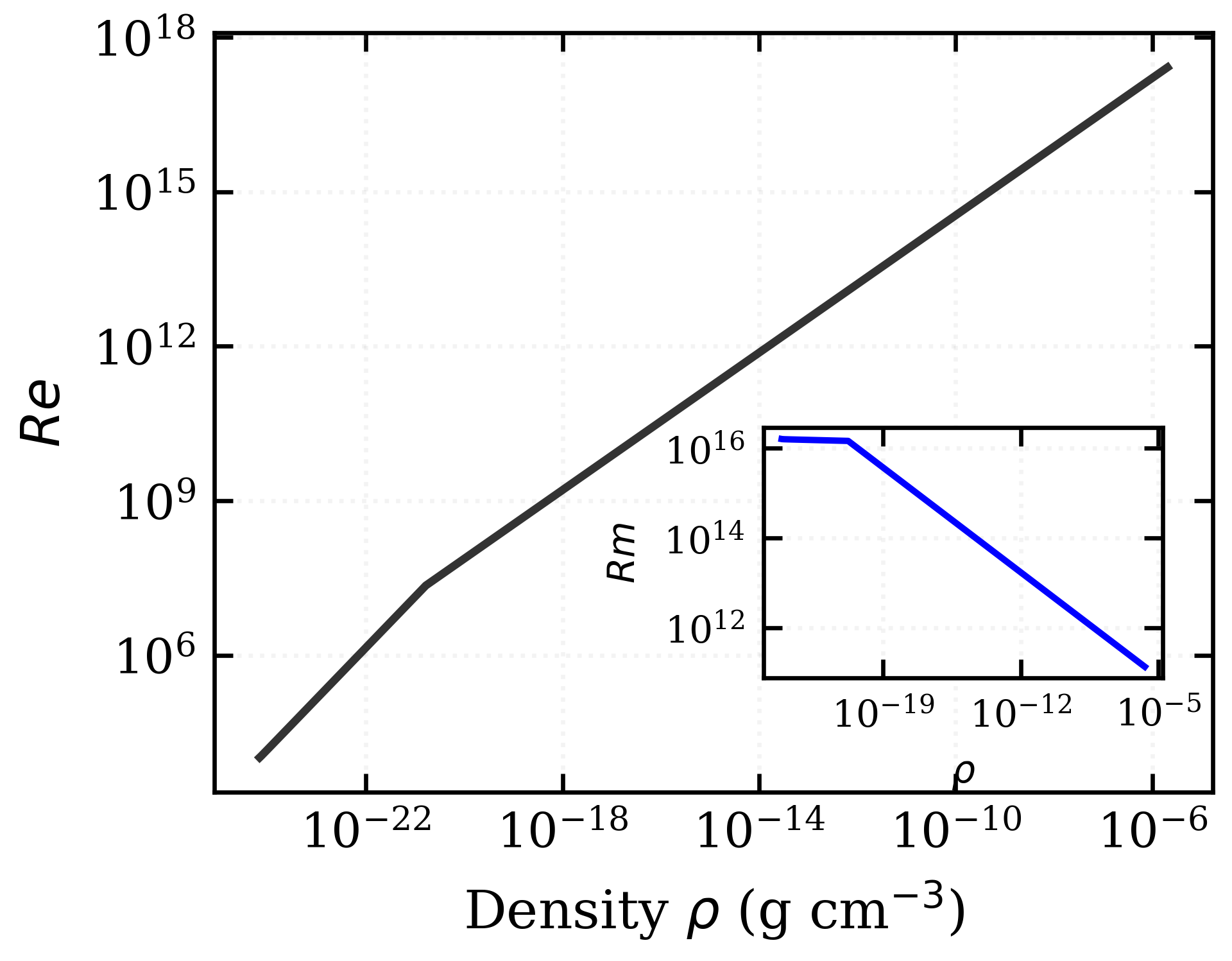}
    \caption{Evolution of $\Rey$ and $\Rm$ (see inset) as a function of density for the fiducial values of first stars given in Table.~\ref{tab:parameters}.}
    \label{fig:ReRm_fidFS}
\end{figure}

Figure~\ref{fig:ReRm_fidFS} shows the evolution of Reynolds numbers in the fiducial first star model. Initially, the magnetic Prandtl number \(\Pm={\Rm}/{\Rey} \sim10^{12}-10^{13},\) is huge. And \(\Rm\) is far above the critical value required for the SSD such that the amplification of magnetic field due to stretching overcomes dissipation. During collapse, \(\Rey\) increases by many orders of magnitude, while \(\Rm\) changes much more gradually. The corresponding decrease in \(\Pm\) eventually brings the first stars into the \(\Pm\lesssim 1\) regime after a density enhancement of \(\sim 10^{12}\). The fiducial first galaxy is followed over a much smaller compression range than the first star case. So its \(\Pm\) evolution is less extreme, nonetheless important for the magnetic field amplification. 

\subsection{Magnetic field evolution}
Turbulent motions stretch, twist and bend an initially weak magnetic field to amplify the magnetic energy until it reaches some fraction of the available turbulent kinetic energy \citep{ruzmaikin1990almighty}. 
From Eq.~\eqref{eq:kcoldynamo3},
the root mean square (rms) magnetic field amplification is
\begin{align}\label{eq:mag1}
    \Brms = B_0\,\xi^{2/3}\, \exp\left({\int \tilde{\gamma}\;{\rm d}\tilde{t}}\right),
\end{align}
where \(B_0\) is the initial seed field, \(\xi^{2/3}\) describes the amplification from flux freezing during spherical collapse, and \(\wt\gamma\) is the supercomoving dynamo growth rate.
In case of $\Pm \gg 1$ systems, the growth rate is set by the viscous eddies with shortest turnover time and hence by $\Rey$. 
Note that the magnetic field growth during collapse is \ti{super-exponential} and faster than the standard exponential growth due to the strengthening of physical turbulent velocity, decrease in length scales, and the rise in $\Rey$ opens a wider inertial range of rapidly stretching eddies. 

Unlike stationary turbulence models, here we deal with the decaying turbulence. Following the treatment of \cite{KS2006Cluster} on the dynamo action of decaying turbulence, we consider the stretching of magnetic field by an eddy of size $\tilde l$, which is supercritical ($\Rm(\tilde{l}) > \Rmcr$) and stays in the inertial range for the relevant time we are interested in. For Kolmogorov turbulence $\wt v_l\propto\wt l^{1/3},$ the growth rate of magnetic field is taken to be 
\begin{align}\label{eq:mag2}
	\wt{\gamma}(\wt{l}) = g \frac{\tilde{v}_l}{\tilde{l}} = g\, \frac{\wt{v}_0}{\wt{l}_0^{1/3}} \frac{1}{\wt{l}^{2/3}},
\end{align}
where $\tilde{l}$ is a fixed scale. The dimensionless growth rate factor \(g\) accounts phenomenologically for the dependence of dynamo growth on compressibility, turbulent Mach number, and partial ionization \citep{Xu_2016, Mckee1}. In particular, numerical simulations show that dynamo growth becomes less efficient as turbulence becomes increasingly compressible and supersonic \citep{2011PhRvL.107k4504F}. We adopt \(g=1\) as the fiducial value, corresponding to a growth rate equal to the turnover rate of the active eddy.

For a fixed active comoving scale \(\wt l\), substitution of Eqs~\eqref{eq:dec2} and \eqref{eq:dec3} into Eq.~\eqref{eq:mag2} gives the kinematic exponent before the physical velocity reaches its cap as
\begin{align}\label{eq:mag3}
    I_1 = \int \tilde\gamma\ \dd \tilde{t} &= \frac{3g}{\alpha \etad}\left(\frac{\tilde{l}_{0i}}{\tilde l}\right)^{2/3}
    \\
    &\lb[\lb(1 + \frac{\etad (1+\alpha)}{2 \lb(\tilde{l}_{0i} / \tilde{v}_{0i}\rb)}\,\tilde{t}\rb)^{2\alpha / 3(1+\alpha)} - 1\rb],\notag
\end{align}
Once the physical turbulent velocity reaches \(\vsat\), the exponent is obtained using Eqs~\eqref{eq:dec2} and \eqref{eq:dec4}
\begin{align}\label{eq:mag4}
   \int \tilde\gamma\ \dd \tilde{t} = I_1 + \frac{g\,\vsat}{\tilde{l}^{2/3}\, \mathcal{L}_{0,\rm sat}^{1/3}} \int_{a_{\vsat}} \tilde{a}\ \dd\tilde{t},
\end{align}
where we evaluate the second term numerically starting from $a_{\vsat}$. 

Closure calculations based on the eddy damped quasi-normal Markovian (EDQNM) approximation show that, for \(\Pm>1\), the dynamo growth rate increases with \(\Rey\) but approaches an asymptotic value above \(\Rey_{\rm asym}\simeq10^6\) \citep{irshad2026smallscaledynamosaturationmagnetic}.
We therefore take the fastest growth rate to be
\begin{equation}\label{eq:gedqnm}
    \Gamma = g\, \frac{v_0}{l_0}\, \Rey_{\rm eff}^{1/2},
\end{equation}
with the effective kinetic Reynolds number \(\Rey_{\rm eff}=\min(\Rey,10^6)\). So as \(\Rey\) grows during collapse,  the growth rate approaches an asymptotic value for \(\Rey > \Rey_{\rm asym}\simeq10^6\). For $\Pm < 1$, relevant during the final stages of first star formation, resistive scales are the fastest and effective. But from \cite{irshad2026smallscaledynamosaturationmagnetic}, when $\Pm < 1$ the growth rate approaches an asymptotic value when $\Rm > \Rm^{\rm asym} \simeq 10^6$. Since $\Rm \gg 10^6$ in first stars at all times, {Eq.~\eqref{eq:gedqnm}} 
works assuming Kolmogorov turbulence is not modified. So this transition in $\Pm$ has no practical consequence for the growth rate in the present model.

In computing \(\Brms\), we first identify the fastest effective eddy available for stretching. During the kinematic stage, this eddy amplifies the field until the magnetic field reaches saturation with that scale
\(\Brms= \sqrt{4\pi f_B\rho}\,v_l.\)
This marks the end of the kinematic stage on that scale. Lorentz force back reaction then suppresses further efficient stretching by the saturated eddy.

Magnetic amplification nevertheless continues because larger eddies in the inertial range contain more kinetic energy, even though their turnover times are longer. We therefore replace the saturated eddy by the next fastest unsaturated eddy and evolve the field using Eq.~\eqref{eq:mag1} until equipartition is reached on that scale as well. Repetition of this process produces a progressive, scale by scale saturation of magnetic energy. We continue this sequence until the magnetic field reaches equipartition with the integral scale.

Once equipartition has reached the largest relevant scales, the magnetic field follows the equipartition \(\Brms \simeq \sqrt{4\pi f_B\rho}\,v_0\). If the physical turbulent velocity is approximately constant, then
\begin{equation}\label{eq:eqscale}
        \Brms\propto\rho^{1/2}.
\end{equation}
If the turbulence continues to evolve during collapse, the saturated field grows more steeply
\begin{equation}\label{eq:ampscale}
    \Brms\propto \rho^{(4+5\alpha)/[6(1+\alpha)]}.
\end{equation}
For \(\alpha=2/3\), this gives \(\Brms\propto\rho^{0.73}\) steeper than the flux freezing scaling \(\Brms\propto\rho^{2/3}\). 

The saturation efficiency \(f_B\) depends on the turbulence properties. Similar to the growth rate factor, it decreases as the flow becomes increasingly compressible and transitions from subsonic to supersonic turbulence \citep{2011PhRvL.107k4504F}. 
Partial ionization can further modify the nonlinear evolution and saturation level, particularly in first-star environments \citep{Xu_2016, 2019MNRAS.487.2673B,Mckee1}.
Corresponding to the maximum saturation efficiency found in the EDQNM simulations we adopt the fiducial value of \(f_B=0.55\) \citep{irshad2026smallscaledynamosaturationmagnetic}.

\subsection{Collapse stopping criterion}
\label{sec:stopping}
The saturated magnetic field strength {can} continue to increase as the gas density rises. We therefore specify a final gas density
\(\rho_{\rm f}\), at which the collapse calculation is terminated.

For first galaxies, the relevant endpoint is the formation of a rotationally supported disc. The angular momentum of a halo is characterized by the dimensionless spin parameter \citep{Mo_book}
\begin{equation}\label{eq:stopcrti1}
\lambda=
\frac{L|E|^{1/2}}{GM^{5/2}},
\end{equation}
where \(L\), \(E\), and \(M\) are the total angular momentum, binding energy, and mass of the halo, respectively. Halo angular momentum is generated primarily by tidal torques from the surrounding large scale structure. Cosmological simulations find an approximately log-normal distribution of initial halo spins, with a typical value \(\lambda_i\simeq0.05\) \citep{2001MNRAS.321..559B, 2013MNRAS.428.2109D, 2014ApJ...781...60H, 2015MNRAS.448..568H}.

We use the disc formation criterion given in \citet{2002thas.book.....P} to estimate the compression factor $\xi = \rho_f / \rho_i$. Let \(M_d\) and \(\lambda_d\) denote the mass and spin parameter of the self gravitating disc at the onset of rotational support. Then the final density of the collapsing gas,{using angular momentum conservation arguments}, is
\begin{equation}\label{eq:stopcrit2}
\rho_{\rm f}^{\rm FG} \simeq 
\rho_i
\left[
\left(\frac{M_d}{M}\right)
\left(\frac{\lambda_d}{\lambda_i}\right)^2
\right]^3,
\end{equation}
where \(\rho_i\) is the initial gas density. We take \(M_d\) to be of order the baryonic mass, \(M_d\simeq(\Omega_b/\Omega_m)M\), and adopt \(\lambda_d\simeq 0.5\) as the criterion for a rotationally supported disc. For this representative halo with \(\lambda_i=0.05\), the density enhancement is \(\xi\simeq 4\times 10^3\). So in the parameter study for FG we vary the initial spin parameter \(\lambda_i\) and determine the final density from Eq.~\eqref{eq:stopcrit2}.

Throughout this work we assume that the angular momentum has a negligible influence on the collapse until the gas approaches centrifugal support. Therefore the collapse is approximated as spherical during the magnetic field amplification, allowing the supercomoving formalism to remain applicable. However, once a rotationally supported disk forms, the assumptions of spherical collapse and isotropic turbulence are expected to break down. Although the subsequent fragmentation and collapse may continue to amplify the magnetic field locally, this phase lies beyond the scope of the present model.

{For first stars, we instead follow the collapse to densities characteristic of protostellar-core formation. The density at which}
this stage is {reached} is determined by the thermochemical evolution of primordial gas, including the cooling and chemical reactions that occur at high density \citep{2009Natur.459...49B,2013ASSL..396..103G}. Since the present model does not follow a chemical reaction network, we select a range of final densities $\rho_{\rm f}^{\rm FS},$ appropriate to the protostellar core formation in the parameter study. 

\section{Results and parameter dependence}
\label{sec:results}

The model {presented in the previous section} tracks the evolution of the initial turbulence and magnetic field during gravitational collapse, using a set of  parameters, that {can be categorized} into three groups: (i) those set by the host dark matter halo and its cosmological context, (ii) those describing the initial turbulence, and (iii) those governing the magnetic field evolution. 
Table~\ref{tab:parameters} summarizes the fiducial values and the explored ranges of the parameters {separately for first stars and first galaxies}. The relevant {model} parameters {describing the cosmological context and host halo} are :

\begin{enumerate}
    \item \ti{Mass of the system} ($M$):
    \begin{itemize}
        \item First stars: We choose  $M=10^6\,M_\odot$, close to the minimum halo mass for efficient H$_2$ cooling \citep{2013ASSL..396..103G}.
        \item First galaxies: We choose  $M=10^8\,M_\odot$, {representative of} atomic-cooling haloes {in which} Ly$\alpha$ line cooling {enables collapse} \citep{1stgalaxies2011}.
    \end{itemize}
    \item \ti{Redshift of formation} ($z$): We {adopt}  $z=20$ for first stars and $z=10$ for first galaxies, typical formation redshifts found in cosmological simulations.
\end{enumerate}

The {parameters governing} turbulence in the system {are:}
\begin{enumerate}
    \item[3.] \ti{Initial integral scale of turbulence:} is parametrized by $f_l = l_{0i}/\rvir$. Simulations of \citet{Higashi_2022} suggest a driving scale of $1/3$ of the Jeans length, while \citet{2020ApJ...903..136G} suggest $1/2$ of the virial radius; we adopt $f_l=1/2$ as the fiducial value.
    \item[4.] \ti{Initial integral scale velocity:} is parametrized by $f_v=v_{0i}/c_s$. \citet{Greif_Firstgalaxy2008} report order-unity turbulent Mach numbers at the virial shock in first galaxies, whereas subsonic turbulence is typical at the corresponding radii in first stars \citep{Greif_2011}. 
    But we adopt a conservative $f_v=0.1$ for both systems
    as fiducial value and larger $f_v$ is explored in the parameter scan.
    \item[5.] \ti{Saturation Mach number:} $\machsat=\vsat/c_s$. \citet{Greif_Firstgalaxy2008} report Mach numbers of $1$--$5$ near the centers of first galaxies, motivating our fiducial choice $\machsat=3$. For first stars, we adopt  $\machsat=1$, based on the transonic velocity dispersions found by \citet{Greif_2011}.
\end{enumerate}

The magnetic field evolution is governed by

\begin{enumerate}
    \item[6.] \ti{Seed field} ($B_0$): Our fiducial value is the Biermann battery field of \Eq{eq:seed}. Additionally we {explore} $B_0$ over a wide range of $10^{-20} - 10^{-10} \G$ {spanning very weak  seeds to substantially stronger initial fields}.
    \item[7.] \ti{Growth rate factor} ($g$): Defined in \Eq{eq:mag2}, sets the efficiency of the kinematic dynamo. Our fiducial choice $g=1$ corresponds to a growth rate set by the eddy turnover rate. We scan $g\in[0.01,1]$ to represent the loss of efficiency in {increasingly} compressible, supersonic turbulence \citep{2011PhRvL.107k4504F}.
    \item[8.] \ti{Saturation efficiency}: Following \citet{irshad2026smallscaledynamosaturationmagnetic} we take $f_B=0.55$ as the fiducial value and scan $f_B\in[0.01,0.55]$.
\end{enumerate}
Finally, we terminate the evolution at
\begin{enumerate}
    \item[9.] \ti{Final gas density} ($n_{\rm f}$): As described in \S\ref{sec:stopping}, for first galaxies the number density $n_{\rm f}^{\rm FG}$ follows from the initial halo spin parameter $\lambda_i$ via the disc-formation criterion. For first stars, no chemical reaction network is included, so we instead select a range for $n_{\rm f}^{\rm FS}$ spanning densities characteristic of protostellar-core formation \citep[see][]{Machida2007, Yoshida_2008}.
\end{enumerate}
In the rest of this section we apply the model to the formation of first stars and first galaxies. The fiducial values and the ranges explored in the parameter scans are summarized in Table~\ref{tab:parameters}.

\begin{table*}[t]
\centering
\caption{Fiducial parameters and ranges adopted for the first star (FS) and first galaxy (FG) calculations. A dash indicates that the {parameter} is not applicable or relevant.}
\label{tab:parameters}
\begin{tabular}{lcccc}
\hline\hline
 Parameters &
\multicolumn{2}{c}{First stars (FS)} &
\multicolumn{2}{c}{First galaxies (FG)} \\
&Fiducial & Range & Fiducial & Range \\
\hline
 $M\,(\Msun)$ &
$10^6$ & $10^5-10^7$ &
$10^8$ & $5\times10^7 - 10^{10}$ \\
$z$ &
20 & 10 - 30 &
10 &  5 - 20 \\

$f_l$ &
$0.5$ & 0.05 - 1 &
$0.5$ & 0.05 - 1 \\

$f_v$ &
0.1 & 0.1 - 0.8 &
0.1 & 0.1 - 1 \\

$\mathcal{M}_{\rm sat}$ &
$1$ & 0.8 - 10 &
$3$ & 1 - 20 \\

$B_0 (\G)$ & See Eq.~\eqref{eq:seed} & $10^{-20} - 10^{-10}$ &
 See Eq.~\eqref{eq:seed} & $10^{-20} - 10^{-10}$ \\

$g$ &
$1$ & $0.01-1$ &
$1$ & $0.01-1$ \\

$f_B$ &
$0.55$ & $0.01-0.55$ &
$0.55$ & $0.01-0.55$ \\

$n^{\rm FS}_{\rm f} (\ccinv)$ &  $10^{18}$  & $10^{15} - 10^{21}$ & -- & -- \\

$\lambda_i$ & --  & -- & 0.05 & 0.005 - 0.1 \\
\hline\hline
\end{tabular}
\end{table*}

\subsection{Fiducial case}
\begin{figure}
    \centering
    \includegraphics[width=\linewidth]{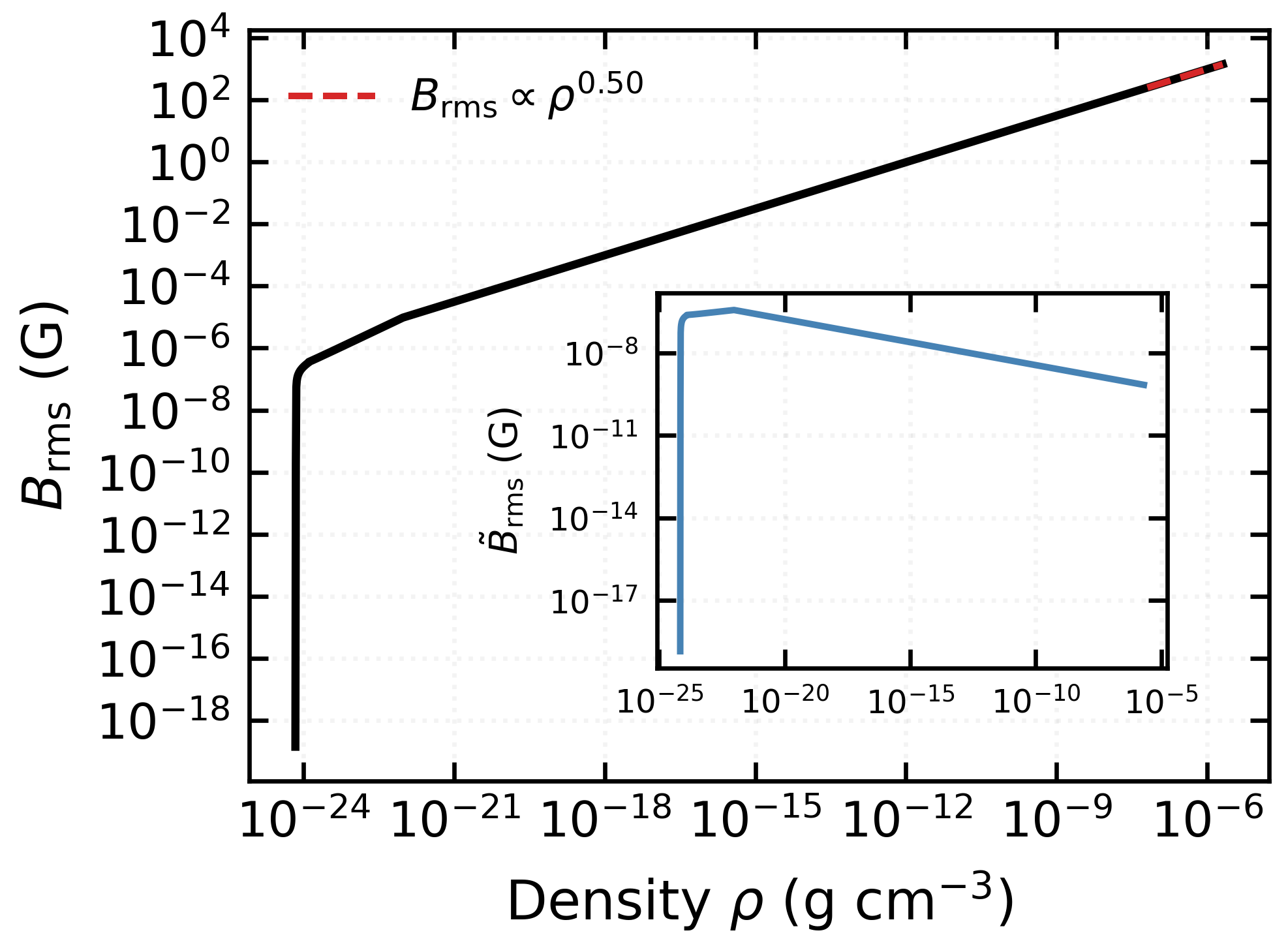}
    \caption{Evolution of the rms magnetic field $\Brms$ (main panel) and the
supercomoving field $\wt{B}_{\rm rms}=(\rho_0/\rho)^{2/3}\Brms$ (inset)
as a function of gas density for the fiducial first stars.
Dashed lines show the best-fit power-law scalings in the nonlinear regime.
Fiducial parameters for each case are listed in \Tab{tab:parameters}.}
    \label{fig:Brms_combined_FS}
\end{figure}

We use the model given in Eqs~\eqref{eq:mag1} - \eqref{eq:mag4} to study the magnetic field evolution during the formation of first stars and first galaxies. {The calculations in this subsection should be regarded as representative realizations of the model rather than unique predictions, since the resulting field strength and correlation scale depend on the uncertain turbulence, dynamo action, and collapse parameters, whose effects are explored in detail in subsequent subsections.} 
\Figs{fig:Brms_combined_FS}{fig:Brms_combined_FG} show $\Brms$ as a function of gas density $\rho$ for the fiducial FS and FG parameter sets (see \Tab{tab:parameters}). Both cases are seeded by the Biermann battery field determined from Eq.~\eqref{eq:seed}. 

The initial kinematic phase of the SSD drives a rapid super-exponential amplification (discussed in detail in the next subsection) of the seed field over many orders of magnitude in both systems,
after which the field transitions into the nonlinear saturation regime.
For FS, spanning eighteen decades of density collapse, the field reaches $\Brms\approx 1.4\,\kG$, in equipartition with outer eddies of size $\approx 13.5\,\AU$.
For FG, which collapses over only three decades,
the final field is $\Brms\approx 75.6\,\mu\G$, in equipartition on a coherence scale $\approx 53.3\,\psc$. {The origin and robustness of these values, and in particular the role played by collapse-induced super-exponential growth, are examined in the following subsections.}

In FS, the turbulent velocity saturates at $\machsat\,c_s$ before the end
of the collapse, so the field tracks equipartition with a constant $v_0 = \vsat$,
giving $\Brms\propto\rho^{0.50}$.
In FG, the velocity is still increasing {as} the field {undergoes} scale by scale {saturation}. Therefore
the nonlinear dynamo receives an additional boost from each newly saturating eddy and increasing $\vrms$, producing a scaling $\Brms\propto\rho^{0.83}$ steeper than
both the flux freezing expectation $\Brms\propto\rho^{2/3}$ and $\Brms\propto\rho^{0.73}$ {from Eq.~\eqref{eq:ampscale} for $\alpha=2/3$} 
expected from equipartition with an amplifying turbulent velocity.
The supercomoving field $\wt{B}_{\rm rms}= \xi^{-2/3}\,\Brms$,
shown in the insets of \Figs{fig:Brms_combined_FS}{fig:Brms_combined_FG}, makes this distinction in scalings clearer. For FS, $\Brms$ grows more slowly than
the flux-freezing, so $\tilde{B}_{\rm rms}$ declines.
Whereas for FG, the field outpaces the amplification due to flux freezing 
throughout the nonlinear stage, confirming the nonlinear dynamo amplification.

{The contrasting FS and FG evolution shows that there is no universal
$B$--$\rho$ scaling in the nonlinear regime: once saturation is complete,
the field can grow more slowly than flux freezing, whereas ongoing
scale-by-scale saturation can drive growth steeper than
$B\propto\rho^{2/3}$. Thus, flux-freezing estimates alone can misrepresent
the magnetic-field evolution during primordial collapse.}
{This is a direct consequence of following the amplification in turbulence and nonlinear stage of the dynamo. Earlier estimates of magnetic field strengths \citep{ScleicherSSD2010, Schober_2012, 2013A&A...560A..87S, Mckee1} due to the dynamo action in collapsing systems
prescribe a fixed turbulent velocity and integral scale. Whereas here they are dynamical variables, evolving under free decay in supercomoving variables while compression amplifies the physical velocity.}

\begin{figure}
    \centering
    \includegraphics[width=\linewidth]{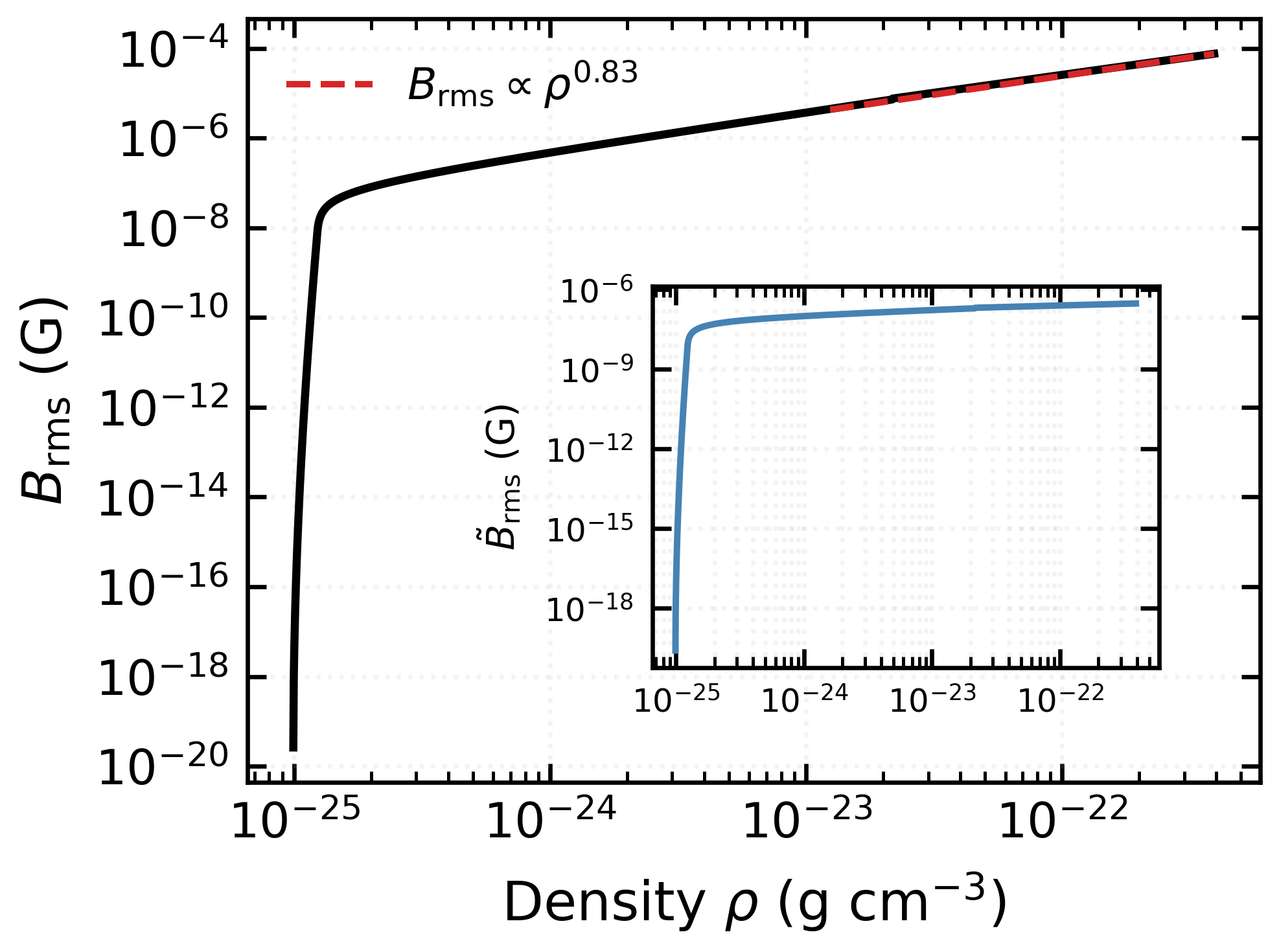}
    \caption{Same as \Fig{fig:Brms_combined_FS} but for first galaxies.}
    \label{fig:Brms_combined_FG}
\end{figure}

\subsection{Super-Exponential Growth and Its Importance}
Standard (stationary background) dynamo amplifies magnetic fields exponentially, with the growth rate proportional to the viscous eddy turn over rate in $\Pm> 1$ systems. For Kolmogorov spectra it is given by $\gamma = v_0\, \Rey^{1/2} / l_0$.
A key feature that distinguishes the turbulent dynamos in collapsing background from its stationary 
counterpart is the faster super-exponential (SE) growth during the kinematic regime over the standard exponential growth. 
This is a result of amplification in the turbulent velocity ($v_0$), and decrease in length scales ($l_0$) {due to} collapse. Moreover, continuous increase in $\Rey$ {with} the increase in density, opens up a wider inertial range and increases $\gamma$. This is important until $\Rey = \Rey_{\rm asym}$ due to the asymptotic growth rate limit found in the EDQNM calculations of \cite{irshad2026smallscaledynamosaturationmagnetic}.

To quantify the speed up resulting from SE growth, we compare the time scales taken by the
collaps{e-driven} and standard dynamos to reach equipartition with the viscous scale eddies of the initial turbulence,
\(\Bnu = \sqrt{4\pi f_B\,\rho_i v_{\nu,i}^2}\). We first evaluate the time
\(\tcol\) for the collapsing dynamo \(\Brms(\tcol) = \Bnu\). 
{To isolate the acceleration of the dynamo from the direct amplification due to compression, we factor out the flux-freezing contribution from the collapsing solution.}
Therefore the comparison {of the standard dynamo} is made against the flux
freezing compensated field rather than \(\Brms\) itself. We evaluate \(\btilde\)
at \(\tcol\) and determine the time \(\tstd\) {required for the standard dynamo to reach the same field strength}. 
Since $\btilde(\tcol) = \Bnu\,\xi(\tcol)^{-2/3} < \Bnu$, the
standard dynamo is still within its kinematic stage. Thus we get \(\tstd = (1 / \gamma_i) \ln{\lb(\btilde(\tcol) / B_0\rb)}\),
where $\gamma_i\equiv\gamma(t=0)$ is determined from the initial values of the
turbulence. The \ti{speed-up} factor is then defined as
\begin{equation}\label{eq:speed}
    S = \frac{\tstd}{\tcol}.
\end{equation}
{Physically, $S$ measures how much faster the dynamo amplifies the field in a
collapsing background than in a static one, after removing the direct
contribution from flux freezing. Thus, $S>1$ indicates a net acceleration
of dynamo growth due to collapse.}
But for the turbulence model we are considering, there is an initial decay in turbulence before compression amplifies it. Due to this for parameters where the standard dynamo is already efficient it is possible to get $S < 1$.

{Previous works have largely focused on conditions favorable to efficient dynamo action. Our fiducial $f_v=0.1$ places the standard dynamo in a comparatively inefficient regime, allowing us to test whether collapse-induced SE growth can compensate under less favorable conditions.} 
But if the dynamo is extremely inefficient, flux freezing dominates even during the kinematic stage. We further explore this trend directly through the growth rate efficiency factor $g$ (Eq.~\eqref{eq:mag2}), sweeping it from its fiducial value $g=1$ down to values as small as $g\sim0.01$, {while keeping} all other parameters {fixed} at their fiducial values (\Tab{tab:parameters}).

To understand this mathematically, let us consider the simpler case of turbulent amplification where $v_0 \propto 1/a$, $l_0 \propto a$ and $\Rey$ is constant. Here the number of e-folds due to the amplification of $\btilde$ and that only due to flux freezing are equal when
\begin{align}
    \ln{\xi^{2/3}} = \tilde\gamma\, \tilde{t},
\end{align}
where $\tilde{\gamma}$ is the initial growth rate and $\tilde t$ is given in Eq.~\eqref{eq:sctime}. So it is clear that for extremely inefficient dynamos ($\tilde\gamma\,\tff \ll 1$), flux freezing is dominant in amplifying the field even during the kinematic stage of SSD. 

\begin{figure}
    \centering
    \includegraphics[width=\linewidth]{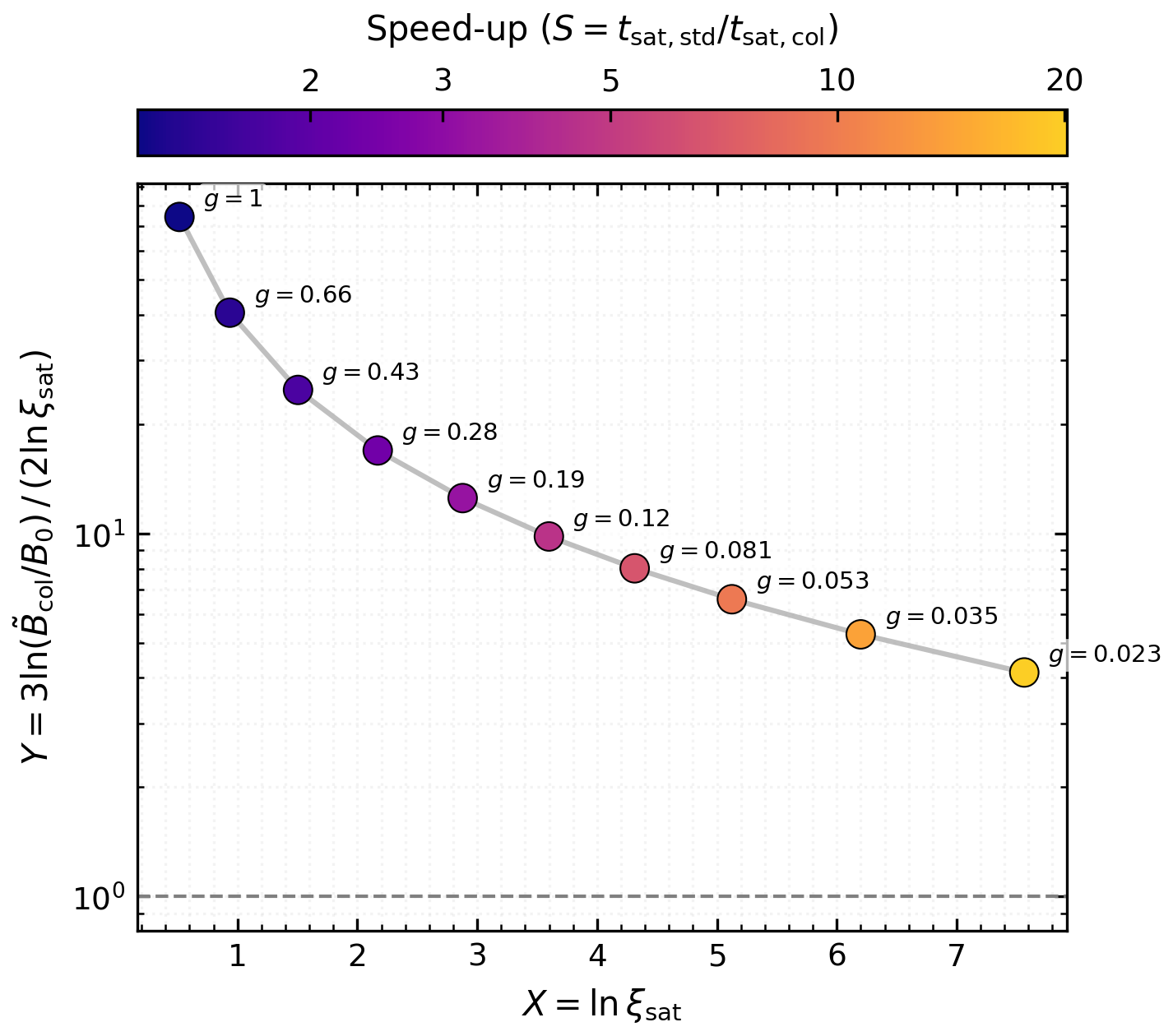}
    \caption{The speed up factor defined in Eq.~\eqref{eq:speed} for the case of FG is shown in the color bar with varying the growth rate factor $g$. 
    For each $g$, the X-axis shows $\ln\xi_{\rm sat}$, the density compression factor at which $\Brms$ reaches viscous equipartition.
    And the Y axis quantifies ratio of e-folding due to $\btilde$ and only flux freezing. $Y = 1$ (dashed line) is shown for reference. 
    }
    \label{fig:SEvsFF_g_FG}
\end{figure}

\begin{figure}
    \centering
    \includegraphics[width=\linewidth]{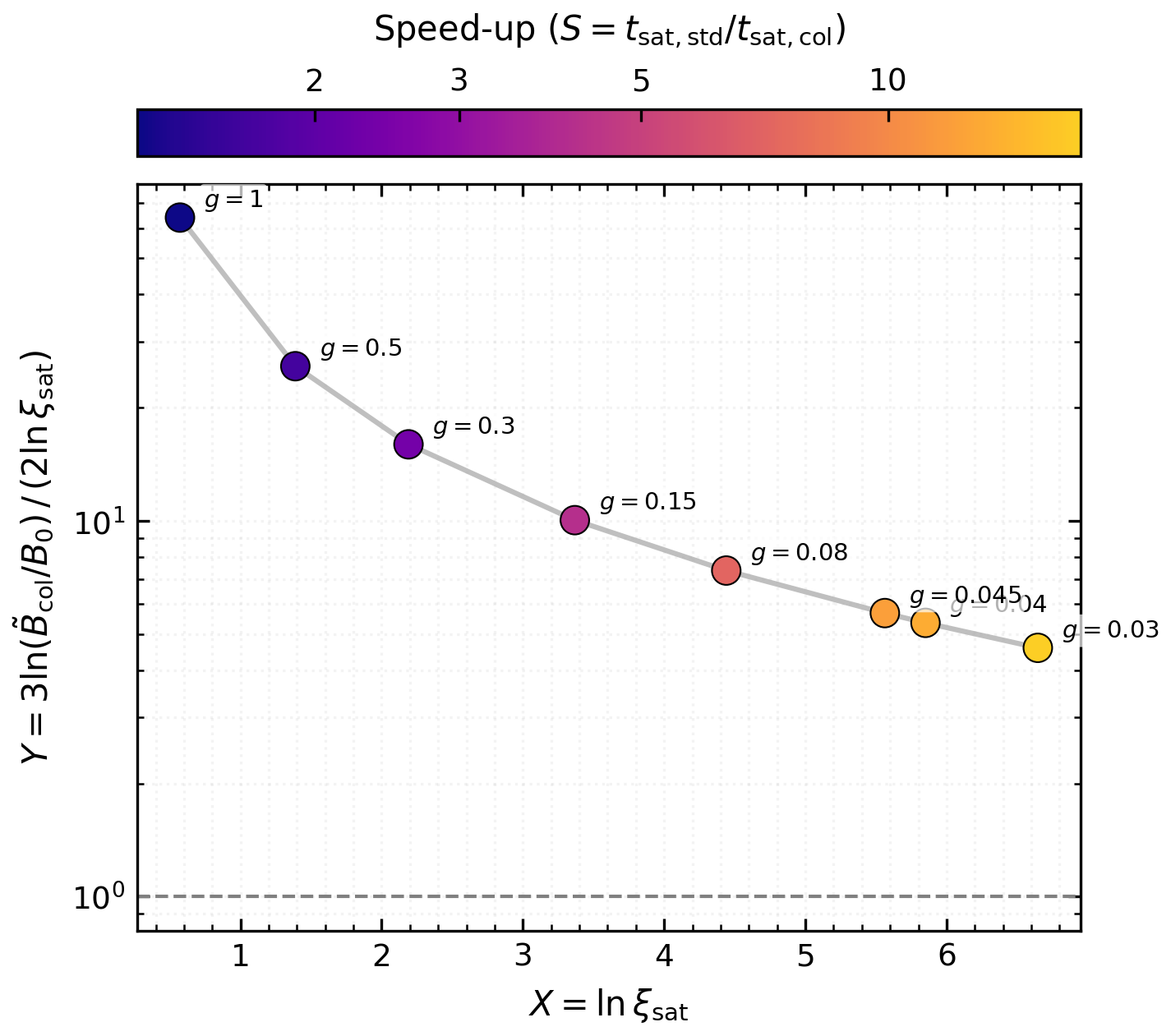}
    \caption{Same as Fig.~\ref{fig:SEvsFF_g_FG} but for First stars.}
    \label{fig:SEvsFF_g_FS}
\end{figure}

The results for FG and FS are summarised in
Figs~\ref{fig:SEvsFF_g_FG} and \ref{fig:SEvsFF_g_FS}
where the horizontal 
axis $X = \ln \xi_{\mathrm{sat}}$ records the density compression factor $\xi_{\rm sat}=\rho(\tcol)/\rho_i$ at the viscous equipartition of collaps{e-accelerated} dynamo $\Brms = \Bnu$. The vertical axis 
$Y = 3\ln(\tilde{B}_{\mathrm{col}}/B_0)\,/\,(2\ln \xi_{\mathrm{sat}})$ 
measures the ratio of {logarithmic amplification produced by the flux-freezing-compensated dynamo} $\btilde$ to that produced due to
compression alone (cf. $\Brms\propto\xi^{2/3}$ in Eq.~\eqref{eq:mag1}). When $Y=1$, flux freezing is as dominant as dynamo. 
And the speed-up factor ($S$) defined in Eq.~\eqref{eq:speed} is shown
in color bar for different growth rate factor $g$.

{Together, \(X\), \(Y\), and \(S\) distinguish the regime in which collapse-induced SE growth is genuinely important. A useful speed-up requires sufficient collapse before saturation, corresponding to larger \(X\), while the dynamo must still dominate over direct flux-freezing amplification, \(Y>1\). We therefore expect little SE enhancement when the dynamo is already very efficient and saturates at small \(X\). The strongest useful enhancement happens at intermediate \(X\) with \(Y>1\). A transition toward flux-freezing-dominated amplification as \(Y\) approaches unity.}

{This expected progression is seen in Figs.~7 and 8 as \(g\) is decreased. At large \(g\), the dynamo saturates after little collapse and \(S\) remains small despite \(Y\gg1\). At intermediate \(g\), saturation occurs later while the dynamo remains dominant over flux freezing, producing the largest useful speed-up. For still smaller \(g\), \(X\) continues to increase but \(Y\) declines toward unity, indicating that the dynamo itself is becoming too inefficient and the system is approaching the flux-freezing-dominated regime.}

For both FS and FG, there is a speed up of $\sim 10$ over the standard dynamo due to the super-exponential growth. 
The speed up is stronger for smaller $g$ due to the longer time available for the SE growth to act. This is crucial
for the quick emergence of strong magnetic field in first structures than expected from the standard dynamo. Note that for the range of $g$ we explored, kinematic dynamo is always dominant over the flux freezing amplification. 

Earlier estimates of dynamo action in collapsing systems have  used stationary growth rates evaluated at fixed turbulent conditions, which amounts to setting $S=1$ by construction, and then asked whether the resulting e-folding time fits within a free-fall time \citep{Schober_2012, 2013A&A...560A..87S, Mckee1}. 
{The collapsing-dynamo framework considered here introduces an additional possibility: even when the standard dynamo is comparatively inefficient, the evolving turbulent conditions during collapse can accelerate the growth sufficiently to achieve $S>1$.}
For the fiducial parameters in Tab.~\ref{tab:parameters} we find $S\approx1$, so the standard dynamo estimate is adequate there. Under the compressible, supersonic conditions expected during structure formation, however, the growth rate factor can be suppressed well below unity \citep{2011PhRvL.107k4504F}, and in that regime the standard dynamo substantially underestimates the field developed within $\tff$.

\begin{figure}
    \centering
    \includegraphics[width=\linewidth]{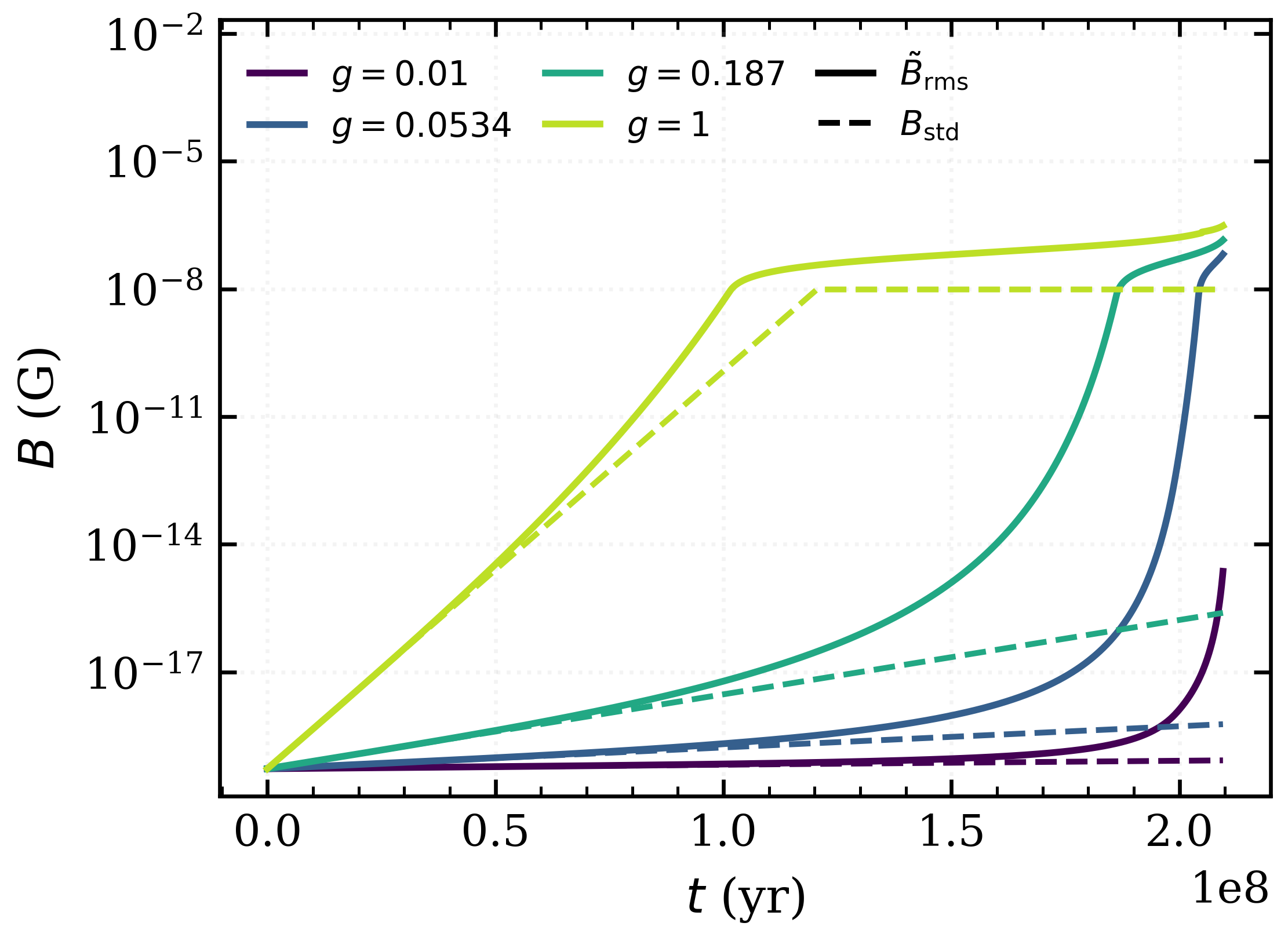}
    \caption{The amplification of flux freezing compensated magnetic field $\btilde$ (solid line) and that due to standard dynamo (dashed line) as a function of time for various growth rate factors $g$ (in color) in the first galaxies.
    }
    \label{fig:SEfield_g_FG}
\end{figure}

\begin{figure}
    \centering
    \includegraphics[width=\linewidth]{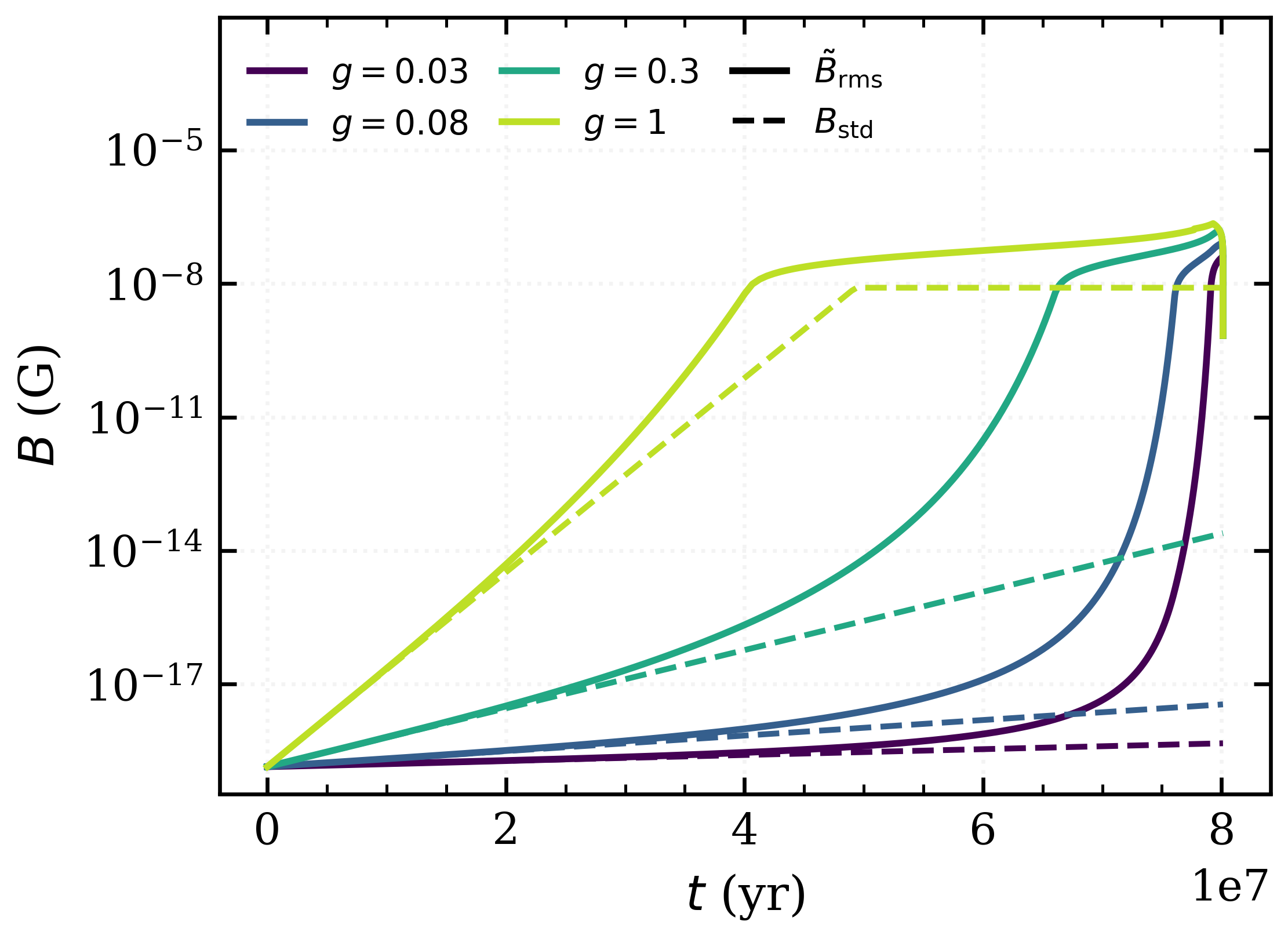}
    \caption{Same as Fig.~\ref{fig:SEfield_g_FG} but for First stars. The reason for final sharp fall is due to the decrease in $\btilde$ we saw in Fig.~\ref{fig:Brms_combined_FS}.}
    \label{fig:SEfield_g_FS}
\end{figure}

Figures~\ref{fig:SEfield_g_FG} and \ref{fig:SEfield_g_FS} make this explicit for FG and FS respectively, showing the flux freezing compensated field $\btilde$ (solid lines) alongside the field of the corresponding standard dynamo (dashed lines) as a function of time, for a range of growth rate factors $g$ (colors). Since flux freezing has been divided out of both, the separation between the solid and dashed curves isolates the dynamo amplification alone, and it already amounts to orders of magnitude in the field strength at the end of the collapse due to the SE growth with speed up of $S\gtrsim 10$.
SE growth also determines whether saturation is reached at all within the available time. For FS the collapsing dynamo saturates for every $g$ considered, whereas for FG the $g=0.01$ case is still in its kinematic stage when the disc forms, the dynamo being too inefficient to saturate within the smaller compression factor in FG.

\subsection{Parameter scan}
\label{sec:paramscan}

\begin{figure*}[t]
    \centering
    \includegraphics[width=\linewidth]{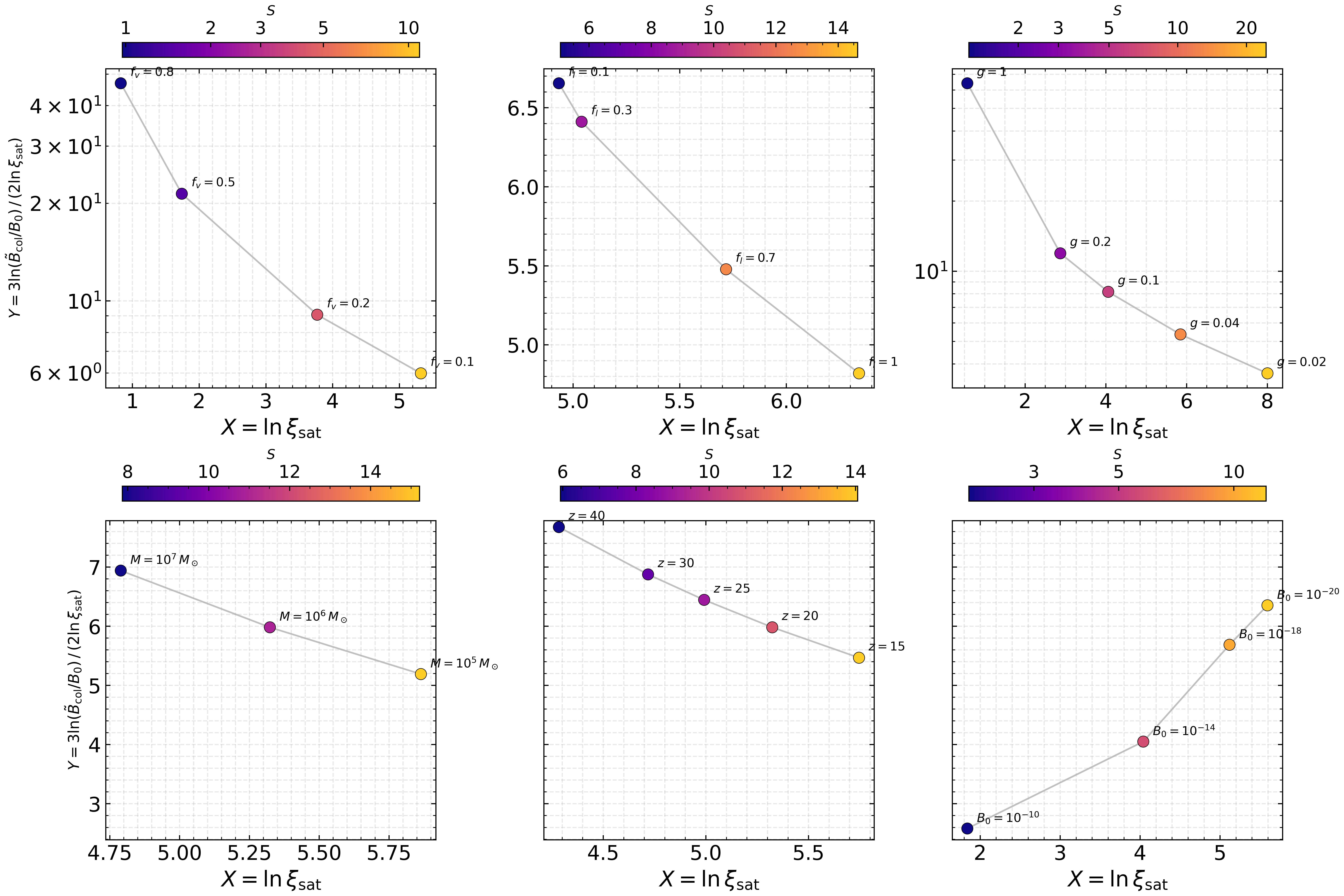}
    \caption{Phase-plane coordinates and speed-up $S$ for First Stars, varying $f_v$, $f_l$, $B_0$, $M$, $z$, and $g$. We take fiducial $g=0.05$ so that the standard dynamo stays inefficient and the speed-up is clearly visible. All other parameters are fixed to fiducial values (Table~\ref{tab:parameters}). Color encodes $S$ within each panel.}
    \label{fig:paramscan_S_FS}
\end{figure*}

\begin{figure*}[t]
    \centering
    \includegraphics[width=\linewidth]{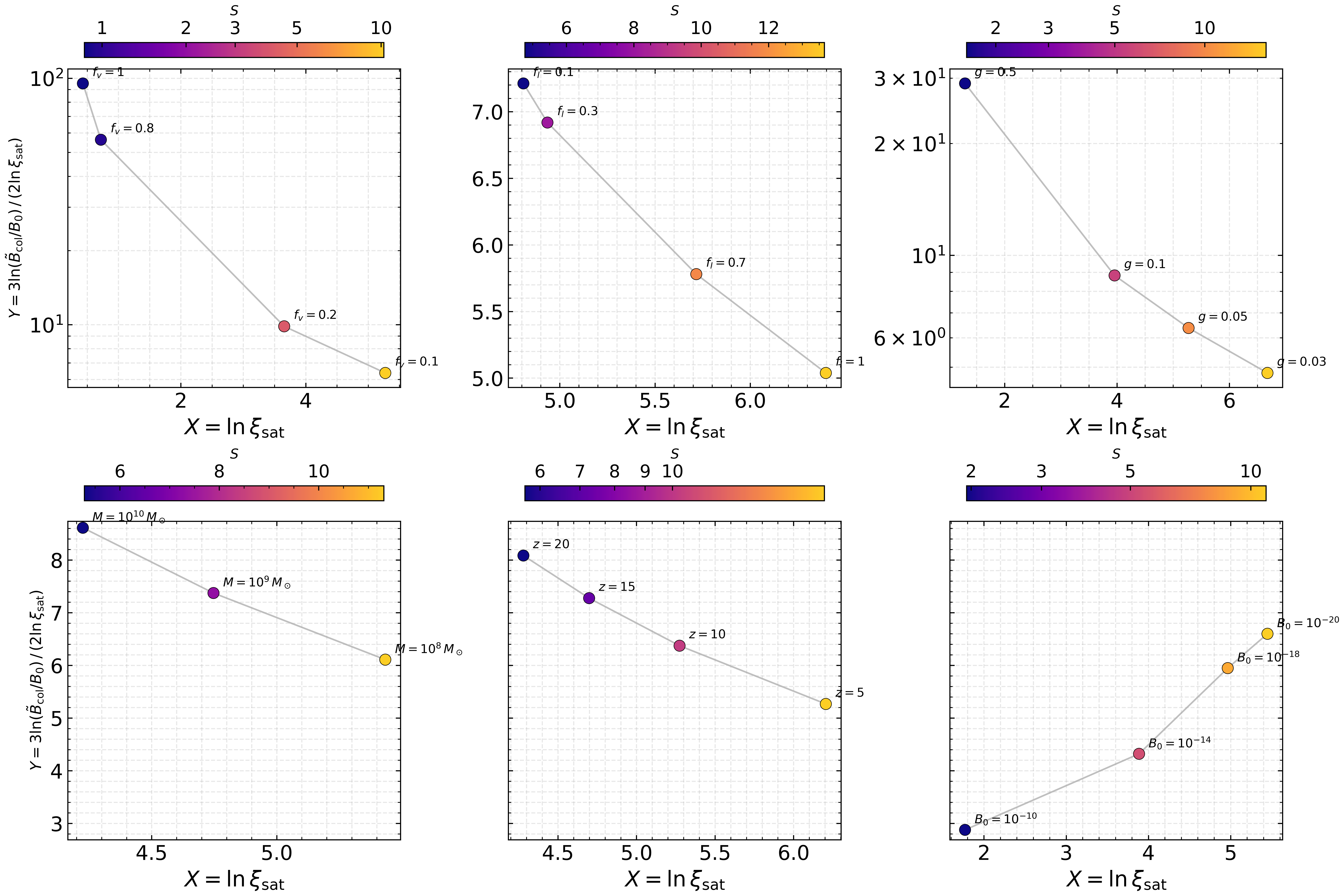}
    \caption{Same as Fig.~\ref{fig:paramscan_S_FS} but for First Galaxies.}
    \label{fig:paramscan_S_FG}
\end{figure*}

\subsubsection{Kinematic stage: Super-exponential growth}
In the last subsection we showed how the speed-up $S$ responds to the growth rate factor $g$. {We now explore its dependence on} 
the remaining parameters. {Unless varied explicitly, we fix \(g=0.05\), that places the system in the regime where SE enhancement is substantial while dynamo amplification still dominates over flux freezing.} All other parameters are held at their fiducial values (see Table~\ref{tab:parameters}).

Figures~\ref{fig:paramscan_S_FS} and \ref{fig:paramscan_S_FG}, for FS and FG respectively, show the variation in speed-up across parameters in the $X$--$Y$ phase plane defined earlier. The trends in $S$ are set by $f_v$, $f_l$, $g$, $M$, $z$, and $B_0$. Since $\lambda_i$ and $\machsat$ matter only for the final saturation, we omit them from the discussion of the kinematic stage.

{Before considering the dependence on each parameter individually, it is useful to identify the basic timescale competition that controls these trends.}
The physical reasoning follows from the time scale ratio $S = \tstd / \tcol$, where $\tstd = \lb(1 / \gamma_i\rb) \ln{\lb(\btilde (\tcol) / B_0\rb)}$, while $\tcol$ itself has no closed form. A longer $\tstd$ implies that the system experiences sufficient collapse such that the turbulence strengthens and the length scales decrease substantially before the dynamo saturates, which in turn shortens $\tcol$. Moreover, {for sufficiently slow dynamos, the collapse time itself limits the duration of amplification, so \(\tcol\) becomes comparable to \(\tff\). We therefore use \(\tcol\sim\tff\) as a useful approximation when interpreting Figs.~\ref{fig:paramscan_S_FS} and \ref{fig:paramscan_S_FG}.} SE growth is therefore important when $\mathcal{T} = \tstd / \tff$ is large, that is, for a longer $\tstd$ (slower standard dynamo) or a shorter $\tff$ (faster collapse).

The mass and formation redshift sets the strength and scale of turbulence through the virial radius $\rvir \propto M^{1/3}/\lb(1 + z\rb)$ and the sound speed $c_s \propto M^{1/3}\sqrt{1+z}$. Since the densities increase as $(1+z)^3$, {and temperature varies as $\Tvir \propto c_s^2 \propto M^{2/3}\, (1+z)$,} the kinematic viscosity of the plasma at the start of the collapse scales as {$\nu \propto n_n^{-1}\, T^{1/2} \propto M^{1/3}(1+z)^{-5/2}$} from Eq.~\eqref{eq:nu}. But note that due to the asymptotic growth rate of the kinematic stage mentioned earlier, the decrease in $\nu$ only makes the dynamo more efficient until $\Rey = \Rey_{\rm asym}$.
The redshift alone sets the free fall time, $\tff \propto \lb(1+z\rb)^{-3/2}$. Thus
\begin{align}\label{eq:Tscaling}
    \mathcal{T} \propto \frac{f_l^{1/2}}{g\lb(1+z\rb)\, M^{1/{6}}\, f_v^{3/2}}\ \ln{\lb(\frac{\Bnu}{B_0}\rb)}.
\end{align}
These scalings provide a natural explanation for the trends seen in Figs~\ref{fig:paramscan_S_FS} and \ref{fig:paramscan_S_FG}.
A decrease in $B_0$, $g$, $z$, $M$ or $f_v$, and an increase in $f_l$, all increase $\mathcal{T}=\tstd/\tff$, {either by slowing the standard dynamo or by increasing the number of e-foldings required before saturation, leading to prominence of SE growth.} 
{These trends are also reflected in the trajectories through the \(X\)--\(Y\) plane. For parameters that reduce the dynamo efficiency, saturation occurs after a greater degree of collapse, shifting the system to larger \(X\), while \(Y\) decreases as dynamo amplification becomes progressively less dominant over flux freezing. The largest useful speed-ups therefore occur at large \(X\) while remaining in the \(Y>1\) regime. The dependence on \(B_0\) is somewhat different: a weaker seed requires more dynamo e-foldings before saturation, shifting the system to both larger \(X\) and larger \(Y\) and thereby increasing the opportunity for SE amplification.}

In every panel the larger speed-ups lie at larger $X=\ln\xi_{\rm sat}$. 
{Over the parameter range explored here, the resulting amplification remains dominated by this collapse-accelerated dynamo growth, with flux freezing providing a subdominant contribution.}
Note that the case of $f_l$ {is more subtle}, if it decreases too much then instead of an efficient dynamo we get an inefficient one, due to the associated faster initial decay in velocity as we saw earlier in Fig.~\ref{fig:vel_fvfl}.

\subsubsection{Saturation: final field vs parameters}
\label{sec:paramscan_finalB}

\begin{figure*}[t]
    \centering
    \includegraphics[width=\linewidth]{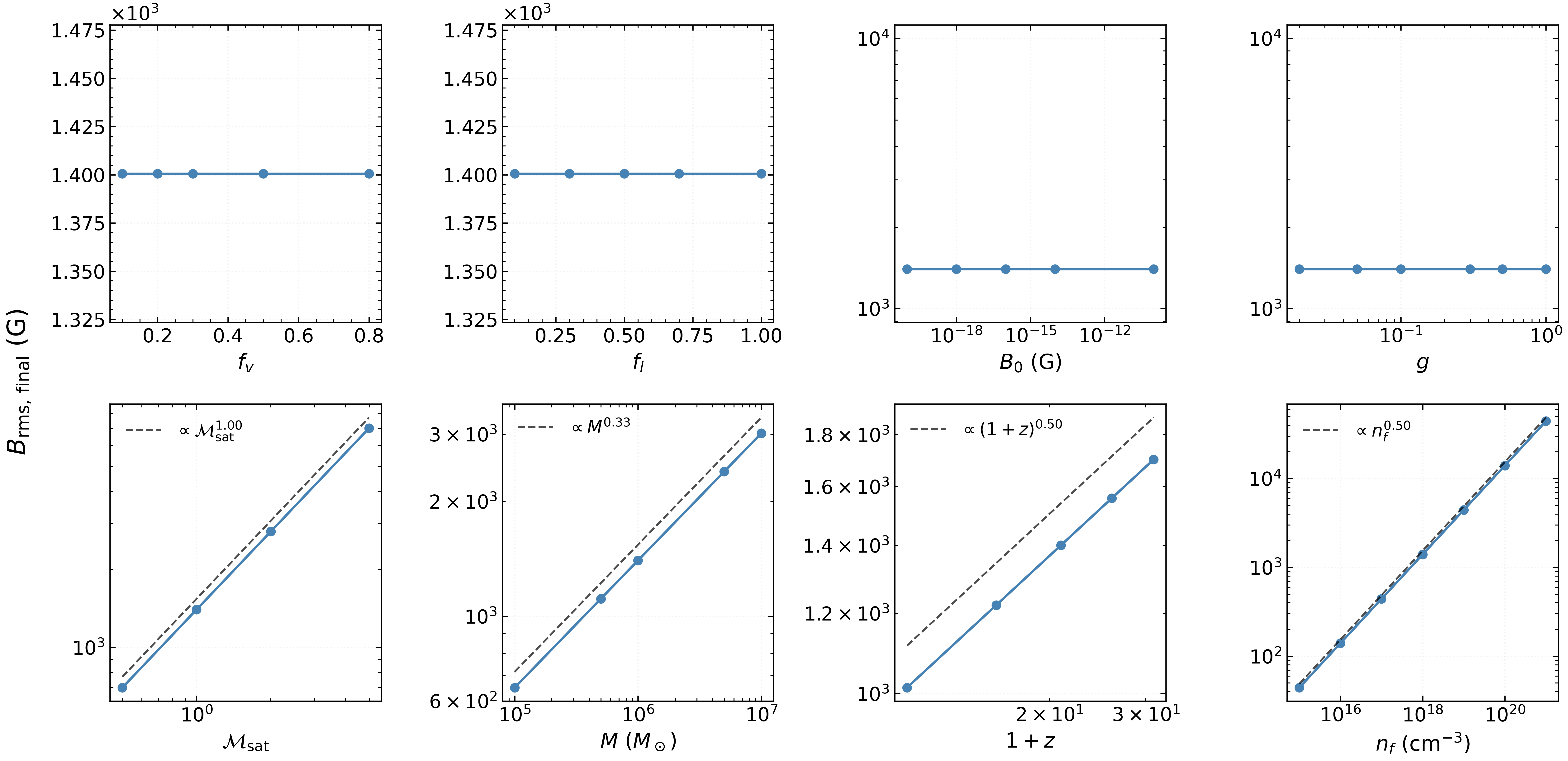}
    \caption{Magnetic field strength at the end of the collapse ($\bfin$) for $g=1$ as a function of different parameters for FS. Dashed lines show the best-fit power law.}
    \label{fig:paramscan_finalB_FS}
\end{figure*}

\begin{figure*}
    \centering
    \includegraphics[width=\linewidth]{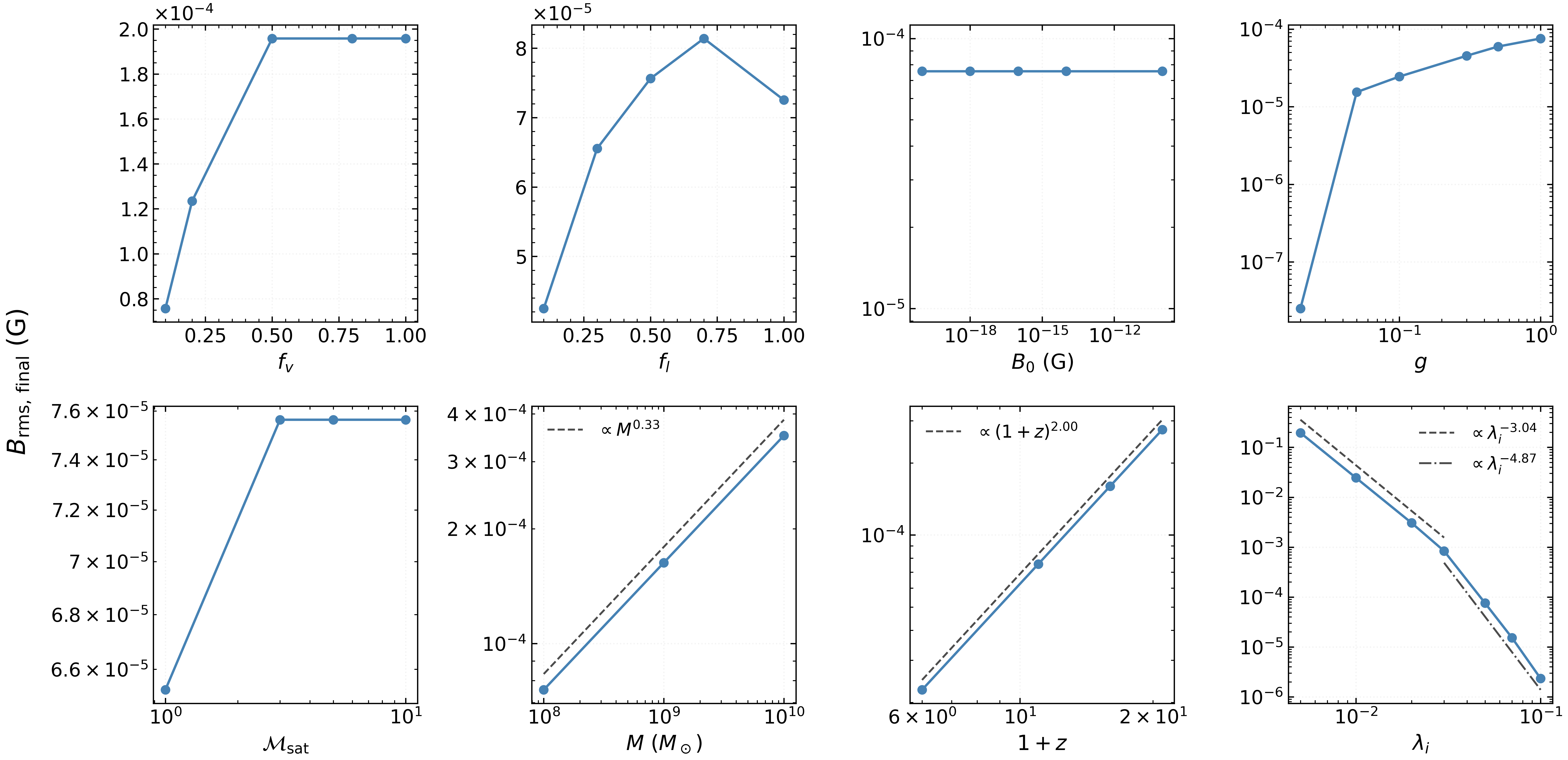}
    \caption{Same as Fig.~\ref{fig:paramscan_finalB_FS} but for First Galaxies. Instead of final number density $n_f,$ the last subplot is for $\bfin$ vs $\spin$, where the fit is done using double power law shown by the dashed and dash - dotted lines. }
    \label{fig:paramscan_finalB_FG}
\end{figure*}

We now turn from the speed-up during the kinematic stage to the final magnetic field strength. \Figs{fig:paramscan_finalB_FS}{fig:paramscan_finalB_FG} show this for the variation of each parameter, keeping all others at their fiducial values. The best-fit power law scalings of the final $\Brms$ as a function of $\machsat$, $M$, $z$ and $n_f$ are shown by the dashed lines.
For $\lambda_i$, a double power law gives a better fit.

In the case of FS, the turbulent velocity always saturates by the end of collapse. So interpreting the trend here is straightforward. Assuming the dynamo to be saturated by the end of the collapse, the final magnetic field strength follows $\bfin = \sqrt{4\pi f_B\,\rho_f}\, \machsat c_s$. Since the endpoint $\rho_f$ is fixed directly by $n_f^{\rm FS}$, the parameters $f_v$, $f_l$, $B_0$ and $g$, which are crucial in the kinematic stage, play no role in setting $\bfin$; accordingly Fig.~\ref{fig:paramscan_finalB_FS} shows flat lines for these.

Meanwhile, we recover the scalings expected from equipartition, $\bfin \propto \machsat$ and $\bfin \propto n_f^{1/2}$. To understand the scaling of $\bfin$ with $M$ and $z$, note that at fixed $n_f$ only $c_s \propto M^{1/3}\,\sqrt{1+z}$ varies, giving $\bfin \propto M^{0.33}$ and $\bfin \propto (1+z)^{1/2}$. 

For FG, {the total compression is smaller than in FS because we terminate the evolution once the gas reaches rotational support and forms a disc.} {At this stage,} the turbulent velocity may not have reached its saturation value. This is clear from Fig.~\ref{fig:paramscan_finalB_FG}, where $\bfin$ is almost independent of $\machsat$ for larger saturation velocities.
{Consequently}, unlike the FS case, {the final field in FG remains sensitive to} $f_v$, $f_l$ and $g$, {since these parameters continue to influence the dynamo before full saturation is reached}. In contrast, $\bfin$ remains independent of $B_0$, confirming that these dynamos have finished their kinematic stage and are in the nonlinear regime. Once the dynamo is efficient enough, at higher $f_v$ and $g$, the dependence flattens out. The dependence on $f_l$ is instead non-monotonic, peaking near $f_l\simeq0.7$: decreasing $f_l$ enhances the initial turbulent decay, while increasing it towards $f_l=1$ lengthens the eddy turnover time so that the dynamo is inefficient.

Nonetheless, {several of the power-law scalings found for FS also persist in FG}, because most of the parameter scans reach equipartition by the end of the collapse and we have $\bfin = \sqrt{4\pi f_B\,\rho_f}\, v_0$, where $v_0$ may not have reached $\vsat$. Note from Eq.~\eqref{eq:stopcrit2} that with $M_d\simeq(\Omega_b/\Omega_m)M$ the compression factor $\xi$ is independent of both $M$ and $z$, so these parameters enter only through $c_s$ and $\rho_i$. Thus, similar to FS, we get $\bfin \propto M^{1/3}$ because $v_0 \propto c_s \propto M^{1/3}$ with increasing mass. In the case of redshift, $c_s \propto \sqrt{1+z}$ and the final density inherits $\rho_i \propto (1 + z)^3$, so that $\bfin \propto (1+z)^2$ follows for FG. {Interestingly, the mass exponent lies within the range $\bar{B}\propto M_{\rm dyn}^{0.25\text{--}0.4}$ inferred for the large-scale magnetic fields of nearby galaxies as a function of their dynamical mass \citep{Tabatabaei2016}. This raises an intriguing possibility that
part of the observed mass dependence is already established when the small-scale dynamo saturates during halo collapse, with subsequent galactic evolution and field ordering modifying, but not necessarily erasing, this scaling.}

Finally we fit $\bfin$ against  $\lambda_i$ through a {broken} power law. From Eq.~\eqref{eq:stopcrit2} we have $\rho_f^{\rm FG} \propto \spin^{-6}$. So when $\spin$ is small, the compression is enormous, such that $v_0$ saturates.
Thus the final magnetic field follows $\bfin \propto \sqrt{\rho_f^{\rm FG}} \propto \spin^{-3}$, and a similar scaling is recovered. In the case of high $\spin$, the velocity is still in its amplification stage. From the asymptotic scaling discussed around Eq.~\eqref{eq:dec3}, we have $v_0 \propto \rho^{7/30}$. Then we get $\bfin \propto \sqrt{\rho v_0^2} \propto \rho^{11/15} \propto \spin^{-4.4}$. The recovered scaling is slightly steeper because, for high $\spin$, {the limited compression prevents the dynamo from reaching full} equipartition and thus the weaker field is still dwelling in the nonlinear stage.

From the parameter scans and scalings we found, we quote the magnetic field strength for FS and FG, assuming a steady state turbulent velocity and a saturated magnetic field, as
\begin{equation}\label{eq:Bfs}
\begin{split}
    B_{\rm FS} \approx 1.4\,\kG\,
    &\lb(\frac{f_B}{0.55}\rb)^{1/2}
     \lb(\frac{n_f}{10^{18}\,\ccinv}\rb)^{1/2} \\
    &\times \lb(\frac{M}{10^6\,\Msun}\rb)^{1/3}
     \lb(\frac{1+z}{21}\rb)^{1/2},
\end{split}
\end{equation}
\begin{equation}
\begin{split}\label{eq:Bfg}
    B_{\rm FG} \approx 196\,\mu{\rm G}\,
    &\lb(\frac{f_B}{0.55}\rb)^{1/2}
     \lb(\frac{\spin}{0.05}\rb)^{-3} \\
    &\times \lb(\frac{M}{10^8\,\Msun}\rb)^{1/3}
     \lb(\frac{1+z}{11}\rb)^{2}.
\end{split}
\end{equation}
where we fixed $\machsat$ to its fiducial value. We note that for the smallest saturation efficiencies, the $\bfin$ quoted here can be smaller by a factor of $5$--$10$. Also, in the case of FG, the field given above should be taken as an upper limit, because the velocity may not have saturated and the dynamo may not reach saturation by the time the disc forms.
{These above expressions should be interpreted as scaling estimates rather than unique predictions: they capture how the final field depends on the dominant astrophysical parameters. The parameter dependences are more robust than the absolute normalization, which remains sensitive to the uncertain saturation physics and collapse endpoint.}

\begin{figure*}[t]
    \centering
    \includegraphics[width=\linewidth]{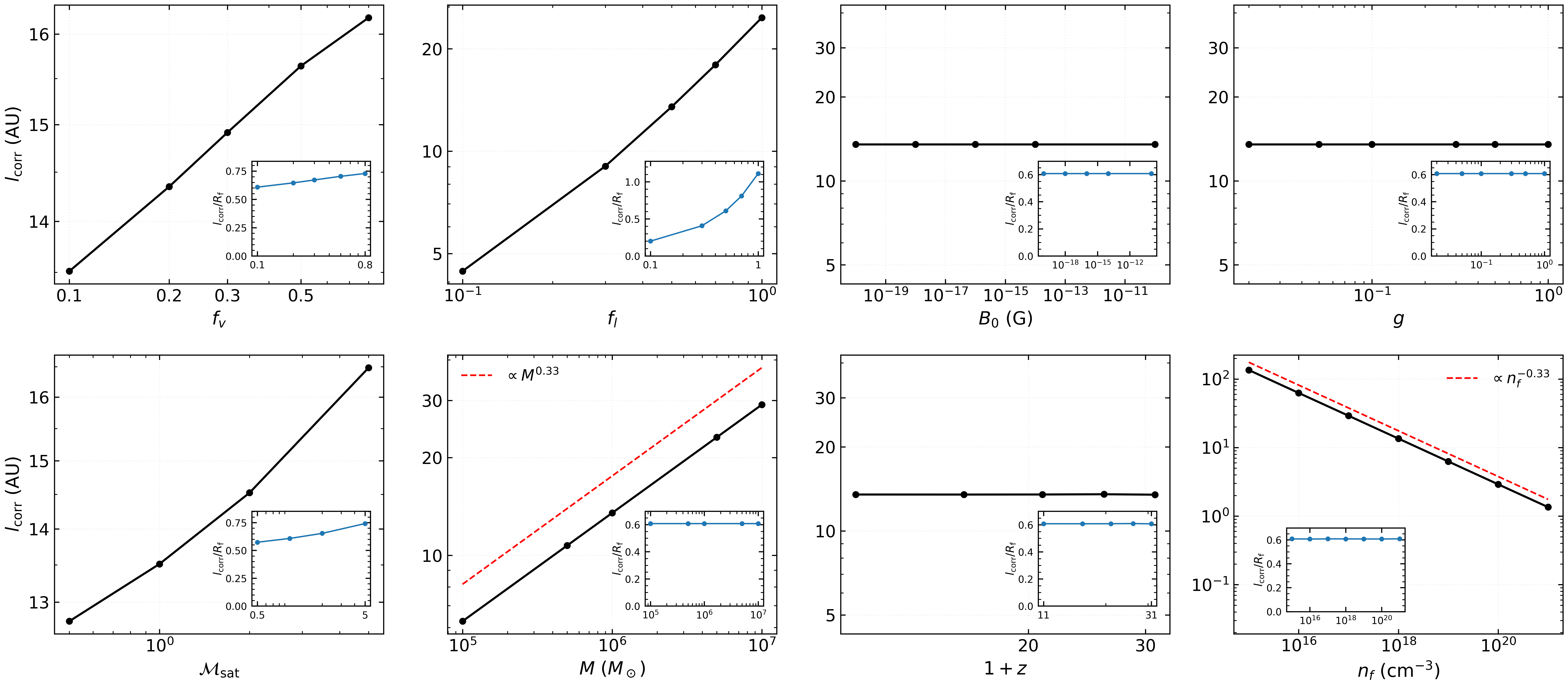}
    \caption{Correlation scale of the magnetic field at the end of the collapse ($\lcorr$) for $g=1$ as a function of different parameters for FS. Dashed lines show the best-fit power law. {Inset focuses on the ratio $\lcorr / R_{\rm f}$, where $R_{\rm f}$ is the final size of the system.}
    }
    \label{fig:paramscan_lcorr_FS}
\end{figure*}

\begin{figure*}
    \centering
    \includegraphics[width=\linewidth]{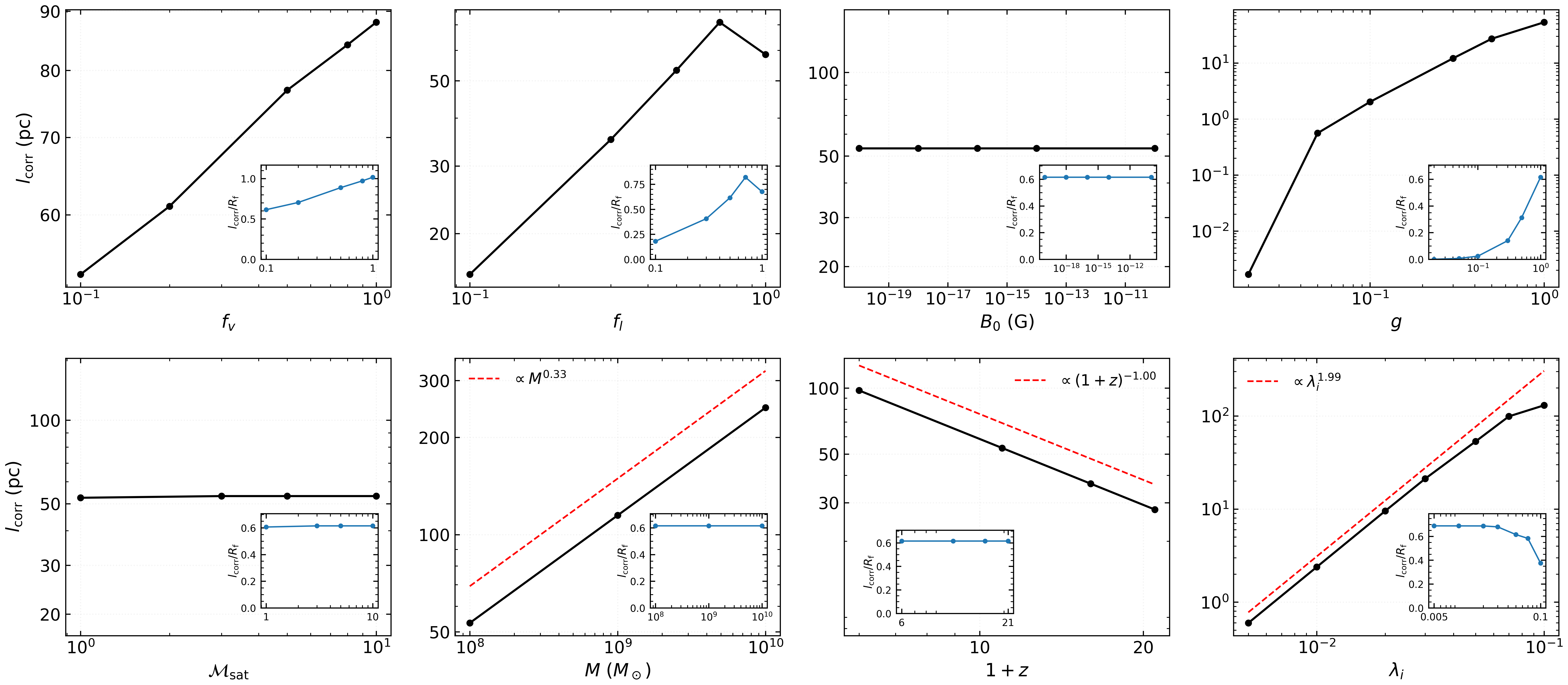}
    \caption{Same as Fig.~\ref{fig:paramscan_lcorr_FS} but for First Galaxies. Instead of final number density $n_f,$ the last subplot is for $\lcorr$ vs $\spin$.}
    \label{fig:paramscan_lcorr_FG}
\end{figure*}

The final correlation scales of the magnetic field, defined as the scale on which the field reaches equipartition, are shown in Figs~\ref{fig:paramscan_lcorr_FS} and \ref{fig:paramscan_lcorr_FG} for FS and FG respectively. 
In both scenarios the final correlation scale of the field, $l_{\rm corr}$, is set by the size of the collapsed system. For FS, the dynamo always reaches saturation and $l_{\rm corr}$ equals the final outer scale of the turbulence for all runs, thus its value is independent of $B_0$, $g$ and is fixed by the size of the core, $R_{\rm f}=a_{\rm f}R_{\rm vir}=[3M_{\rm b}/(4\pi\rho_{\rm f})]^{1/3}$. This gives $l_{\rm corr}\propto M^{1/3}n_{\rm f}^{-1/3}$, with typical values of  $\approx 13.5\AU$ 
at the fiducial $n_{\rm f}=10^{18}\,{\rm cm^{-3}}$. The expected $R_{\rm vir}\propto(1+z)^{-1}$ dependence cancels because the FS collapse ends at a fixed final density, $a_{\rm f}\propto(1+z)$ compensates it.
{The inset plots focus on the ratio of correlation scale}
to the system size. At the fiducial values {$l_{\rm corr}/R_{\rm f}\simeq 0.6$} and changes only through the parameters that set the turbulent outer scale. For FG, the field typically reaches a correlation scale of $\sim50$\,pc, again a sizable fraction ($\simeq0.6$) of the final system size. Since the FG collapse ends at an $a_{\rm f}$ set by Eq.~\eqref{eq:stopcrit2}, we get $l_{\rm corr}\propto\lambda_i^{2}$, and from the $\rvir$ scaling, we have $l_{\rm corr}\propto M^{1/3}(1+z)^{-1}$. Unlike FS, however, FG does not always saturate at the outer scale. For parameters with small growth rates, the dynamo is still in its kinematic stage when the collapse ends and $l_{\rm corr}$ {is much smaller}. Overall, whenever the dynamo saturates before the end of the collapse, the magnetic field ends up coherent on a scale comparable to the system itself, ${\sim}0.6\,R_{\rm f}$. 
{For some extreme parameter choices, the model yields \(l_{\rm corr}/R_{\rm f}>1\), which is clearly not physical. This occurs because the turbulent outer scale is evolved independently of the system size: while compression reduces the physical scale through the factor \(a\), turbulent decay simultaneously shifts the surviving turbulent energy toward larger supercomoving scales, causing \(l_0/R\) to increase. In this regime the simple decay prescription should be regarded as having reached its limit of applicability, with the outer scale physically bounded by the system size.}
{Note that} from SSD simulations and closure models, the peak of the magnetic energy spectra is {expected} to lie between $\sim 1/10$ and $\sim 1/3$ of the forcing scale \citep{Bhat2014,  irshad2026smallscaledynamosaturationmagnetic}.
{This difference arises because our model tracks the largest turbulent scale that has reached equipartition: through scale-by-scale saturation this can extend up to the forcing scale, whereas the peak of the magnetic-energy spectrum remains at a smaller fraction of that scale.}
So {more realistically} we expect a typical magnetic correlation scale of $\sim 6~$--$~20\,\%$ of the system scale for the dynamos in saturation by the end of the collapse.

\section{Impact of the small-scale dynamo generated fields}
\label{sec:impact}

{Throughout the preceding sections, we have followed magnetic-field amplification due to the small-scale dynamo during gravitational collapse. This is motivated by an important difference between the first cosmic structures and}
present day star forming environments. {Simulations of the latter} begin with a coherent field,
because molecular clouds are threaded by the large scale galactic field
maintained by a mean-field dynamo. No such mean field is typically expected in
minihaloes and in atomic cooling haloes at high redshifts where the first stars and galaxies going to form.
{A coherent primordial field could in principle provide a large-scale component, but here we assume it to be sufficiently weak that the SSD-amplified field dominates the magnetic dynamics.}
In this section we investigate the impact of the magnetic fields amplified by SSD during the formation of first stars and galaxies,
on the thermodynamics of the gas, structure formation and astrophysical feedback. 

\subsection{Plasma beta}
{The relevance of the fields obtained above is determined not simply by their absolute strength, but by their pressure relative to the gas, quantified by the plasma beta \(\beta=P_{\rm th}/P_B\). Values \(\beta\gg1\) correspond to dynamically weak fields, whereas \(\beta\sim1\) implies that magnetic forces can influence the collapse. Our model therefore allows us to ask whether the SSD-generated fields become dynamically important before the end of the collapse.}
Assuming a simplified equation of state $P_{\rm th}=\rho c_s^2$ during collapse, we get
\begin{equation}\label{eq:beta}
    \beta = \frac{P_{\rm th}}{P_B} = \frac{8\pi\rho c_s^2}{B^2}.
\end{equation}
The evolution of $\beta$ during the collapse follows from the
$B$--$\rho$ scaling, assuming negligible variation in $c_s$. For $\Brms\propto\rho^{b}$, we get
$\beta\propto\rho^{1-2b}$. So pure flux freezing ($b=2/3$) gives
$\beta\propto\rho^{-1/3}$, while equipartition with a constant turbulent
velocity ($b=1/2$) leaves $\beta$ unchanged. For the steeper scalings we predicted in Eq.~\eqref{eq:ampscale}, due to enhancement in turbulent velocity during collapse, we get 
\begin{equation}
    \beta \propto \rho^{-\lb({1+2\alpha}\rb)/\lb[3(1+\alpha)\rb]}.
\end{equation}
For $\alpha = 2/3$ considered here, $\beta\propto \rho^{-7/15}$ follows a steeper variation than expected from flux freezing alone.

For FS the turbulent velocity reaches its cap before the end
of the collapse and the magnetic field also saturates at \(B^2 = 4\pi f_B\rho\machsat^2 c_s^2\). So for the fiducial values of $\machsat\simeq 1$, $f_B = 0.55$, we get a constant $\beta\simeq 3.64$.
Similarly, for the fiducial $\machsat=3$ in FG, we get $\beta\simeq0.4$ for the field strength given in Eq.~\eqref{eq:Bfg}. Since the latter is upper limit on the field strength, we regard $\beta\simeq0.4$ as a lower limit, rather than as a prediction of
magnetic domination.

{The main implication is that SSD amplification during collapse can bring the system to \(\beta\) of order unity, so that magnetic forces can become dynamically important even much before the end of the collapse.
These values motivate the questions considered below: whether such fields can alter fragmentation and angular-momentum transport, and whether their finite correlation scale is sufficient to affect outflows.}

\subsection{Fragmentation and initial mass function}
The simplest way the field can influence fragmentation is by providing additional pressure support of $P_{\rm th} / \beta$ against gravity. 
The gas then behaves as if it had a larger effective sound speed of
$c_{\rm s, eff}^2=c_s^2\,(1+1/\beta)$. The corresponding Jeans mass which scales as
$M_{\rm J}\propto c_{s,\rm eff}^3\ \rho^{-1/2}$ {is increased} by a factor $(1+1/\beta)^{3/2}$. For $\beta_{\rm FS} \simeq 3.64,$ and the lower limit $\beta_{\rm FG} \simeq 0.4$ quoted in the previous subsection, the Jeans mass increases by only $\simeq1.4$ for FS but $\lesssim 6.5$ for FG. 

{Besides the additional magnetic pressure at the end of collapse, an important aspect is the stage at which dynamically significant fields are established. Collapse-enhanced dynamo growth can bring the field to near-equipartition before the collapse terminates, even when the corresponding stationary dynamo would be comparatively slow. In the FS models, saturation is reached across a broad range of seed strengths and dynamo efficiencies, making the magnetic state of the protostellar gas relatively insensitive to these uncertain initial parameters. In FG, by contrast, saturation before disc formation is conditional on the turbulent and dynamo parameters and on the amount of compression, implying a wider range of magnetic influence on subsequent disc fragmentation. Moreover, nonlinear scale-by-scale saturation produces fields coherent on a non-negligible fraction of the system size, making magnetic braking and angular-momentum transport potentially more important than the modest change in Jeans mass alone \citep{Machida2013, Sadanari2021}.}

{Although previous studies have focused mainly on magnetic braking and angular-momentum transport, some have examined the role of magnetic pressure support in suppressing fragmentation \citep{PriceBate2007, Machida2025}.}
{\citet{HiranoMachida2022} find that an amplified field can suppress disc fragmentation around Pop~III protostars, while \citet{Stacy2022} obtain a single $\simeq30\,\Msun$ protostar where the corresponding unmagnetized calculation fragments into several objects. \citet{Sharda2025} likewise find that magnetic fields modify mass transport through the disc, reducing the mass of the most massive star and shifting the resulting stellar-mass distribution. Taken together, these studies suggest that dynamically significant fields of the kind obtained in our calculations can reduce fragmentation and multiplicity, suppress the formation of low-mass Pop~III stars, and thereby shift the primordial mass function toward higher masses {with the highest mass limited by the magnetic fields through regulating the mass transport,} although the quantitative outcome remains simulation-dependent.}

Gravitational compression also raises the turbulence that drives the
dynamo, and turbulence acts on fragmentation in the opposite sense to the
magnetic field. \citet{Clark2011ApJ} find that primordial gas fragments
readily even for subsonic turbulence, with rms velocities as low as
$0.2\,c_s$, producing fragment masses spread over more than two decades
and hence a broad mass function. \citet{Wollenberg2020} similarly find
that increasing the turbulent energy promotes fragmentation and flattens
the protostellar mass function, whereas the magnetic field suppresses
fragmentation and drives it towards higher masses.

The two effects cannot be varied independently in a collapsing cloud. The
saturated field is tied to the turbulent velocity through equipartition. Hence the net effect on the primordial initial mass
function follows from the competition between the two.

\subsection{Correlation scale of the field and outflows}
{In a stationary SSD, scale-by-scale saturation proceeds relatively slowly through progressively larger eddies. During collapse, the enhanced nonlinear growth accelerates this progression, allowing the field to reach equipartition with much larger eddies within a free-fall time and thereby increasing the magnetic correlation scale to a substantial fraction of the system size.
As stated before,}
by the end of the fiducial collapse of FS and FG, the
field is in equipartition {with the flow} out to $\sim6$--$20\,\%$ of the instantaneous system scale
(\S\ref{sec:paramscan_finalB}). So the SSD generated fields therefore become {considerably more coherent}. 

The correlation scale of magnetic field is directly relevant to the launching of protostellar
outflows. Simulations of \citet{Machida2006} find that an initial ordered field drives strong
jets, whereas \citet{Gerrard2019}
find that a completely tangled field launches no jet at all. We argue that the fields generated due to the nonlinear small scale dynamo
lie much closer to the former case than to the latter, thus we expect outflows from the first structures.
But exactly how the
outflow efficiency changes as a function of the
correlation length is unclear. {This provides a natural target for simulations initialized with saturated SSD magnetic spectra having different peak scales, which could determine the degree of magnetic coherence required for launching efficient outflows.}

\subsection{Non-ideal MHD effects}
One of the crucial non-ideal MHD processes during the star formation is the
ambipolar diffusion that arises from the drift of the ions relative to the neutrals due to the Lorentz force. 
{We consider two consequences of ambipolar diffusion for the fields obtained above, in this subsection: whether it can dissipate the dynamo-generated field during collapse, and whether the associated ion--neutral drift can heat the gas strongly enough to modify the thermal evolution.}
Associated ambipolar diffusivity is given by (\S2.6.5 of
\citealt{SS21})
\begin{equation}\label{eq:etaAD}
    \eta_{\rm AD}=\frac{\Brms^2}{4\pi\gamma_{\rm AD}\,\rho_i\,\rho_n},
    \qquad
    \gamma_{\rm AD}=\frac{\langle\sigma v\rangle_{\rm in}}{m_i+m_n},
\end{equation}
where $\rho_i\simeq x_e\rho$, {up to the ion-to-mean-particle mass ratio, assuming quasi-neutrality} and $\rho_n=(1-x_e)\rho$ are the ion and neutral
mass densities, $\langle\sigma v\rangle_{\rm in}$ is the ion--neutral
momentum transfer rate coefficient {with reaction cross section $\sigma$ and relative velocity $v$,} 
$m_i$, $m_n$ are the ion and
neutral masses \citep{Draine1986}. Unlike the {microphysical} magnetic diffusivity $\eta$, $\eta_{\rm AD}$ depends
on the field itself. 
A field correlated on $\lcorr$ diffuses away in a time
$\lcorr^2/(\eta+\eta_{\rm AD})$. Since the dynamo operates on the
free-fall time, those fields survive as long as
\begin{equation}\label{eq:nonideal}
    \frac{\lcorr^2}{\eta+\eta_{\rm AD}} > \tff.
\end{equation}

Ambipolar diffusion also heats the gas, at a rate
$\Gamma_{\rm AD}\simeq\eta_{\rm AD}B_\ell^2/(4\pi\ell^2)$ \citep{SS21}. Here $\eta_{\rm AD}$ carries the total field, since the
drift is driven by the Lorentz force, while the 
gradient is set by the smallest scale on which the field varies. 
The heating rate scales as
$\Gamma_{\rm AD}\propto\Brms^4 x_e^{-1}\ell^{-2}$, so it is acutely
sensitive to both the ionization history and the scale on which the field
varies. 

The strong ambipolar heating reported by \citet{Schleicher2009}
arises at densities where the proton abundance drops and Li$^+$
becomes the dominant charge carrier, where $x_e = 10^{-10}$ differing by six orders of
magnitude from the fiducial $x_e$ we considered. This reduction would make the
heating dynamically relevant. Whereas 
\citet{Nakauchi2019} find $x_e$ to be two to three orders of magnitude
higher than previously computed at $10^{14}$--$10^{18}\,\ccinv$, and
\citet{Sadanari2022}, whose non-ideal MHD simulations reach protostellar
densities with dynamo amplified fields of $10^{3}$--$10^{5}\,\G$
comparable to ours, find the ambipolar heating rate to stay below the
compressional rate throughout. 

Crucially, none of these calculations
include super-exponential growth {considered here.} 
{By bringing the magnetic field to saturation at lower densities and producing steeper \(B\)--\(\rho\) scalings than a stationary-dynamo estimate, this can enhance ambipolar heating precisely over the density range where the ionization fraction is declining, before three-body \(\mathrm{H}_2\) formation becomes the dominant heating process. If sufficiently strong, this earlier heating could modify the thermal trajectory of the collapsing gas, {increasing accretion rate onto the protostar,} increasing thermal support and potentially suppressing fragmentation. At the same time, ambipolar diffusion dissipates magnetic structure and reduces magnetic support, so its net effect on the collapse cannot be inferred from the heating rate alone.}
Establishing
whether ambipolar heating becomes competitive there requires non-ideal
calculations that follow the chemistry alongside a collapsing dynamo, and
we leave this to future work.

\section{Summary, Conclusions and discussions}
\label{sec:conclusions}

Turbulent dynamos {amplify and} maintain astrophysical magnetic fields by converting 
turbulent kinetic energy to magnetic energy. The kinematic dynamo stage in 
a stationary background exponentially amplifies the initial seed field with the spectral
peak at the resistive scales. Once the 
field is strong enough, Lorentz force back-reacts on the flow, and dynamo enters
nonlinear regime and the exponential growth {transitions} to a linear growth of magnetic
energy in case of Kolmogorov turbulence. During this, the magnetic field can become
ordered through the scale by scale saturation with the kinetic energy. 

Above picture gets modified when we consider the turbulent dynamo
during the formation of stars and galaxies. During gravitational collapse, length scales
decrease and turbulent velocity increases. Earlier in
\cite{irshad}, we developed a framework to analyse dynamo action in collapsing clouds
based on the supercomoving formulation of MHD proposed by \cite{shapiro97}. We showed
that the kinematic dynamo amplifies the magnetic field super-exponentially instead of exponentially, 
due to the increasing eddy turn over rate during collapse. At saturation, the magnetic field
follows a steeper scaling with density than expected from pure flux freezing, even in the unforced case. 
{When the system is unforced}, the initial turbulence decays through the turbulent cascade, however it also amplifies due to  
the gravitational compression. In this work, we extended the analysis to the nonlinear dynamo stage
connecting kinematic and saturation regimes. We showed that the collapse induced boost gives a 
super-linear growth of magnetic energy, instead of linear growth for Kolmogorov turbulence, thus 
boosting the slowest stage of the small scale dynamo.

Based on the supercomoving turbulent decay considered in the supplementary material of \cite{irshad}, 
here we have developed a model for the SSD action during the formation of the first stars in minihaloes 
and galaxies in atomic cooling haloes. Overall the magnetic field evolution is governed by the following parameters: (i) those set by the host dark matter halo, mass and redshift, (ii) those describing the initial turbulence, such as initial velocity and length scale, and (iii) those governing the magnetic field evolution includes seed field, growth rate factor, and 
saturation efficiency. The calculations are only terminated at the protostellar densities for first stars and for first galaxies, a rotationally supported disc formation based on its initial spin is used as the criterion. 
The seed field and initial turbulence are assumed to get generated at the virial shocks.

Countering the turbulent dissipation, gravitational compression amplifies turbulent velocities during
collapse. We cap the turbulent Mach numbers ($\machsat$), as they  approach $\machsat\gtrsim 1$,  
typically reported in the simulations of first stars and first galaxies.
Unlike earlier models of dynamo action during structure formation, here we follow the 
amplification of turbulence and super-exponential growth of magnetic field. 
In the kinematic stage, {we believe that the fastest available eddies dominate the stretching until the magnetic field reaches equipartition with their kinetic energy}. And we follow the nonlinear stage by considering the amplification due to larger and slower eddies in the inertial range, as the fields undergo scale by scale saturation. Finally, at saturation
the field strength is obtained by considering equipartition with the turbulent kinetic energy. 

We quantify the speed up due to the SE growth by taking the ratio between the time taken by the  dynamo in stationary background to reach
equipartition with the viscous scale energy and that taken by the dynamo during collapse. 
The importance of this speed-up, however, depends on the
efficiency of the underlying dynamo. If the standard dynamo is already very efficient to reach saturation well before a free-fall time,
the speed-up {is close to} unity. At the other extreme, if the dynamo is too
inefficient, magnetic amplification becomes increasingly dominated by
flux freezing {and the field remains below saturation}, {preventing it from becoming dynamically important before the collapse terminates}. Therefore
SE growth is important  in the intermediate regime, where
the stationary dynamo is slow enough for substantial collapse to occur,
but remains efficient enough for dynamo amplification to dominate over
flux freezing. 

The parameter study shows the importance of SE growth with speed-up reaching  $\gtrsim 10$ 
for smaller initial turbulent velocities, weaker seed fields, smaller
growth-rate factors, and longer initial turbulent scales. 
Halo mass and formation redshift modify the characteristic sound speed, density, and free-fall time, and
therefore affect the amount of collapse available for dynamo amplification.
These trends demonstrate that the acceleration is not a single universal
multiplicative correction to the stationary dynamo; rather, it emerges
from the coupled evolution of the turbulence and the collapsing medium.
So SE growth due to dynamos is crucial for the accurate prediction of magnetic fields at the end of the collapse {and, as argued in our earlier work \citep{irshad}, provides a natural route toward explaining the rapid emergence of strong magnetic fields in high-redshift galaxies}.

Our model also tracks the nonlinear and saturation stages to estimate the magnetic
field strength and correlation scale {from the scale-by-scale progression of equipartition to larger eddies}. In the FS case, {the large compression drives the turbulent velocity to its saturation value and the magnetic field to equipartition, after which the field follows the saturated scaling \(B\propto\rho^{1/2}\).} While for FG, due to the smaller compression {given} with the onset of rotational support, velocity and magnetic field may not reach saturation leading to steeper $B$--$\rho$ scaling than flux-freezing. 
The parameter scans provide useful scaling relations for these final
fields from eqs~\eqref{eq:Bfs} and~\eqref{eq:Bfg}. {The fiducial field strengths are not unique predictions, whereas the parameter dependence provide more general scaling relations.} Their main utility is
to provide a compact description of how the field strength responds to
the halo and collapse parameters, and to provide physically motivated
initial conditions for simulations that follow the subsequent stages of
star or galaxy formation.

{The dynamically significant fields produced in our model can influence fragmentation through both magnetic support and angular-momentum transport, as found in simulations of primordial star formation \citep{HiranoMachida2022, Sharda2025}. Importantly, collapse-induced SE growth allows such fields to be established earlier than a stationary-dynamo estimate would suggest, extending the density range over which magnetic effects can operate. At the same time, compression also amplifies the turbulence, which tends to promote fragmentation \citep{Clark2011ApJ, Wollenberg2020}. The resulting fragmentation therefore reflects the coupled competition between turbulence and the magnetic field generated by it.}

An important consequence of the nonlinear evolution is that the magnetic
field is not expected to remain concentrated exclusively at the smallest
dissipative scales. In our model, scale-by-scale saturation allows the
magnetic field to approach equipartition on scales extending to a
substantial fraction of the final system size. For the fiducial
first-star and first-galaxy models, we infer the correlation scale of the equipartition
field to be $\sim 6~-~20\% $ of the final system size. The precise
value is uncertain because it depends on the detailed magnetic spectrum,
which is not explicitly evolved in our present model. 

A sufficiently coherent magnetic field can facilitate the launching of magnetic outflows \citep{Machida2006, Gerrard2019}. 
The SSD generated fields can be correlated approximately on $6-20\%$ of the system size. Whether fields with 
such correlation scales can drive outflows and determining how the strength and efficiency of outflows depend on the
magnetic field geometries are still not clear. Another possibility to generate coherent
fields on scales of the system size during the formation of FS and FG is an 
$\alpha^2$-dynamo considered in \citet{irshad}. This requires kinetic helicity, which is expected in such rotating
systems. Thus in future, measurement of the kinetic helicity in simulations with self gravity {will help to assess large-scale field generation \citep{Liao2021}.}

The model also has important limitations. The collapse is assumed to be
homologous and spherical, with rotation neglected until centrifugal
support is reached. But this approximation is useful for following the main
amplification phase.
The thermodynamic treatment is similarly simplified. We adopt an
approximately isothermal evolution at the virial temperature and do not
follow the chemical and radiative processes that determine the detailed
thermal evolution of primordial gas. 
The dynamo framework is derived assuming a highly
conducting plasma, while the gas relevant for primordial star
formation can have a very small ionization fraction. We have only included the impact of
neutrals through viscosity and effective dynamo parameters. Ideally, one {may} need to
solve the two-fluid equations governing both neutrals and ions. 

The weak coupling between ions and neutrals leads to ambipolar diffusion. It can dissipate magnetic structure and can heat the plasma.
Previous calculations have reached different conclusions about
the importance of ambipolar heating in primordial collapse, in part
because the ionization history and the magnetic-field evolution differ
between models \citep{Schleicher2009, Nakauchi2019, Sadanari2022}. The
stronger amplification and steeper scalings found in our collapse-induced dynamo
model makes this question particularly relevant. A self-consistent
calculation combining the collapse boosted dynamo, non-ideal MHD, and primordial chemistry
is therefore needed to determine whether ambipolar heating can
significantly modify the thermal evolution.

\section*{Acknowledgements}
We thank Kandaswamy Subramanian for useful insights and stimulating discussions during the Physics of Galaxy Formation School organized at the Inter-University Centre for Astronomy and Astrophysics (IUCAA). We thank Enrique V\'azquez-Semadeni for helpful interactions during the MIST2025 -- \ti{Cosmic Turbulence and Magnetic Fields: Physics of Baryonic Matter Across Time and Scales} and the \ti{International Symposium on Cosmic Magnetic Fields}. MIP acknowledges useful interactions with Mordecai Mac Low and sincerely thanks Edith Falgarone and JinLin Han for their support during the conferences.

This research was supported by the Department of Atomic Energy, Government of India, under Project No.~RTI4019. We acknowledge the use of artificial intelligence tools in assisting with the development of the Python scripts used for the model presented in this work.

\bibliography{reference}

\end{document}